\documentclass{aa}
\pdfoutput=1
\usepackage{graphicx}
\usepackage{txfonts}
\usepackage{longtable}
\begin{document}

\title{A UV and optical study of 18 old novae with Gaia DR2 distances:
mass accretion rates, physical parameters, and MMRD}

\author{Pierluigi Selvelli \inst{1} \and Roberto Gilmozzi \inst{2} }
\institute{INAF - Ossservatorio Astronomico di
  Trieste, Via Tiepolo 11,
  I -34143 Trieste, Italy\\
  \email{selvelli@oats.inaf.it}
\and European Southern Observatory,
  Karl Schwarzschild-Str. 2, 85748 Garching,
  Germany\\
  \email{rgilmozz@eso.org}
\thanks{ Based mainly on INES data from the
IUE satellite. Other UV data utilized in this paper were obtained
from the Multimission Archive at the Space Telescope Science
Institute (MAST), see Paper I.
This work has made use of data from the European Space Agency (ESA)
mission {\it Gaia} (https://www.cosmos.esa.int/gaia), processed by
the {\it Gaia} Data Processing and Analysis Consortium (DPAC,
https://www.cosmos.esa.int/web/gaia/dpac/consortium). Funding
for the DPAC has been provided by national institutions, in particular
the institutions participating in the {\it Gaia} Multilateral Agreement.
    } }

\date{Received ... ; accepted ....}


\abstract{We combine the results of our earlier study of the UV
  characteristics of 18 classical novae (CNe) with data from the
  literature and with the recent precise distance determinations
  from the Gaia satellite to investigate the statistical properties of
  old novae. All final parameters for the sample include  a
    detailed treatment of the errors and their propagation.  The
  physical properties reported here include the absolute magnitudes at
  maximum and minimum, a new maximum magnitude versus rate of decline (MMRD)
  relation, and the inclination-corrected 1100--6000-\AA\
  accretion disk luminosity. Most importantly, these data have allowed
  us to derive a homogenous set of accretion rates in quiescence
  for the 18 novae.

\smallskip

All novae in the sample were super-Eddington during outburst, with
an average absolute magnitude at maximum of $-7.5\pm1.0$. The average
absolute magnitude at minimum corrected for inclination is
$3.9\pm1.0$.  The median mass accretion rate is
$\log\dot{M}_{1M\odot}=-8.52$ (using $1M\odot$ as WD mass for all
novae) or $\log\dot{M}_{M_{WD}}=-8.48$ (using the individual WD
masses). These values are lower than those assumed in studies of CNe
evolution and appear to attenuate the need for a hibernation
hypothesis to interpret the nova phenomenon.

\smallskip

We identified a number of correlations among the physical parameters
of the quiescent and eruptive phases, some already known but others
new and even surprising. Several quantities correlate with the speed
class $t_3$ including, unexpectedly, the mass accretion rate
($\dot{M)}$. This rate correlates also with the absolute magnitude at minimum
corrected for inclination, and with the outburst amplitude, providing
new and simple ways to estimate $\dot{M}$  through its
functional dependence on (more) easily observed quantities.

\smallskip
There is no correlation between $\dot{M}$ and the orbital period.
}


\titlerunning{Physical parameters and mass accretion rates of old
  novae, after Gaia}
\authorrunning{Selvelli and Gilmozzi}

\maketitle

 %

\section{Introduction}

Old novae are the quiescent stage of systems that have undergone
  a historical classical nova explosion.  Classical novae (CNe) are
  members of the class of cataclysmic variables (CVs), that is, close
  binary systems in which a white dwarf (WD) accretes hydrogen-rich
  material through an accretion disk from a low-mass, near-main-sequence
  companion.  The ``classical nova'' phenomenon is a
  thermonuclear runaway (TNR) event that occurs when the accreted mass
  on the surface of the WD is large enough for the pressure at the
  base of the semi-degenerate shell to initiate nuclear reactions; see
  Starrfield et al. (1985), Shara (1989), Starrfield et al. in Bode
  \& Evans (2008),  Jose' (2016) for comprehensive reviews.

  Theoretical models for the outburst (hereafter OB)
  of CNe (see Shara et al. 1986,
  Livio 1992a, Prialnik \& Kovetz 1992) require low mass accretion
  rates (hereafter $\dot{M}$) during quiescence ($10^{-11}$ to $10^{-9}$
  M$\odot/yr$).  Only at these rates does the material at the base of the
  H-rich envelope remain degenerate enough to ensure the observed
  strong ``flash''. Instead, the available observations of old novae
  indicate higher $\dot{M}$, of order $10^{-9}$ to $10^{-8}$
  M$\odot/yr$.  This apparent disagreement represents a disturbing
  problem in our understanding of the classical nova phenomenon
  (Prialnik \& Kovetz, 1992).

  $\dot{M}$ is a fundamental parameter for our
  understanding of the evolution of CVs in general because these
  systems evolve under the effect of mass transfer.  Angular momentum
  losses are required for mass transfer to occur at all (Hameury
  1994). The standard paradigm of CV evolution assumes that angular
  momentum losses are driven by mechanisms such as magnetic wind
  braking (Verbunt and Zwaan, 1981), dominating in systems with
  $P_{orb} > 3$ hr, and gravitational radiation (Paczynski, 1967, King
  1988), dominating in systems with $P_{orb} < 3$ hr, see also Spruit
  \& Ritter (1983), Howell et al. (1997), Howell et al. (2001), Knigge
  (2006) and Kigge et al. (2011).

  A major problem in the theory comes from the large spread (by at
  least one order of magnitude) in the mass transfer rates at a given
  orbital period (Patterson 1984; Warner 1995; Kolb 2001, Spruit \&
  Taam 2001; Woudt et al 2012). Explanations for this discrepancy
  requiring intermittent cycles produced by nova explosions (Shara et
  al 1986) or by irradiation or mass loss effects (King et al 1996)
  have been proposed.

It should be stressed, however, that reliable statements about  $\dot{M}$ --- the {\it non plus ultra} of binary
evolution according to Patterson (2011) --- can be found only for
objects with well-determined distance and reddening, and by
observations covering the satellite UV range (hereafter, simply UV) because
the bulk of the accretion luminosity is emitted in this spectral
region (Wade \& Hubeny 1998).  The homogeneous UV data on spectral energy distributions (SEDs) in
Paper I are therefore of high relevance for this topic, and since old
novae are (unlike other CVs, e.g., DNe) nearly stable accretors
  (Honeycutt et al 1998; Retter \& Naylor 2000; Puebla et al 2007)
their accretion luminosity and $\dot{M}$ can be determined
with higher precision.

We note that throughout this paper \emph{accretion rate} is used to
  indicate the mass transfer rate through the disk (the one that we
  derive from the observed disk luminosity). Other kinds of accretion
  rates that may be at play in these systems, for example the mass-loss rate
  from the donor, the accretion rate on to the WD, or the mass loss
  from the disk (e.g., via a wind or outflow through the boundary
  layer) cannot be determined from our data, and should not be
  confused with the accretion through the disk.

In Paper I (Selvelli and Gilmozzi, 2013), we studied all available UV
spectra of old novae. This provided  a homogeneous determination
  of several characteristics (and their errors) of the 18 objects in
  the sample: the reddening E(B-V), the SED (well described by a power law), and the total UV-integrated
flux. Paper I concluded that the UV SED is associated to radiation
emitted by an accretion disk, that the disk accounts for most of the
observed UV and optical flux, and that the contributions by the WD and
the cool companion are of second order. The data also indicated
that the quiescent state of classical novae is characterized by a
nearly constant UV luminosity over a time interval of decades.

In this paper we combine the results from Paper I with data in the
literature to derive physical parameters, revisit some known relations
(e.g., the MMRD) and determine precise values of for  $\dot{M}$  during the quiescence phase.

The second release of the Gaia data  (Gaia Collaboration et
  al. 2018) occurred while we were finalizing this paper and we
decided to use the Gaia distances instead of the ones from HST/FGS or
other parallax determinations (e.g., expansion) used earlier. The new
distances mostly fall within the errors of the old ones, but with
superior precision. We have propagated the new input to all
derived physical quantities.

\section{The   basic parameters   and their  uncertainties }

\begin{table*}
\caption{Data. Values in italics for the WD masses in column 13
    indicate the calibrators used in addition to V1500 Cyg.}
\begin{center}
\renewcommand{\tabcolsep}{0.08cm}
\renewcommand{\arraystretch}{1.2}
\begin{tabular}{lrrrrrrrrrrrr}
\hline
\smallskip

& \multicolumn{1}{c}{(2)} & \multicolumn{1}{c}{(3)} & \multicolumn{1}{c}{(4)} &
\multicolumn{1}{c}{(5)} & \multicolumn{1}{c}{(6)} & \multicolumn{1}{c}{(7)} & \multicolumn{1}{c}{(8)} &
\multicolumn{1}{c}{(9)} & \multicolumn{1}{c}{(10)} & \multicolumn{1}{c}{(11)}& \multicolumn{1}{c}{(12)}& \multicolumn{1}{c}{(13)} \\

 Object & \multicolumn{1}{c}{\bf P} & \multicolumn{1}{c}{\bf  $\Delta T$} &\multicolumn{1}{c}{\bf dist} &
 \multicolumn{1}{c}{ $m$v$_{max}$}  & \multicolumn{1}{c}{ $m$v$_{min}$ } &
\multicolumn{1}{c}{\bf  t$_3$} & \multicolumn{1}{c}{\bf t$_2$} & \multicolumn{1}{c}{\bf  A$_V$ }
& \multicolumn{1}{c}{\bf  incl }  & \multicolumn{1}{c}{$\int_{1250}^{3050}
F_{\rm IUE}$} &
\multicolumn{1}{c}{$\int_{1100}^{6000} F_{\rm PL}$} & \multicolumn{1}{c}{$M_{WD}$} \\

& \multicolumn{1}{c}{[d]} & \multicolumn{1}{c}{[yr]} & \multicolumn{1}{c}{[pc]} & \multicolumn{1}{c}{[mag]} & \multicolumn{1}{c}{[mag]} & \multicolumn{1}{c}{[d]} &
\multicolumn{1}{c}{[d]} & \multicolumn{1}{c}{[mag]} &
\multicolumn{1}{c}{[rad]} & \multicolumn{2}{c}{[ $10^{-11}$ erg cm$^{-2}$
s$^{-1}$ ]} & \multicolumn{1}{c}{[M$\odot$]} \\

\hline

V603 Aql &  0.138 &  72 &  311$\pm$7 & $-1.4\pm$0.2 & 11.7$\pm$0.1 &   9$\pm$2 &   5$\pm$3 & 0.25$\pm$0.06 & 0.26$\pm$0.05     &  174$\pm$25.4 &  270$\pm$38  & {\it 1.15$\pm$0.15}      \\
T Aur    &  0.204 &  97 &  857$\pm$38 &  4.2$\pm$0.1 & 15.2$\pm$0.3 & 100$\pm$5 &  80$\pm$3 &  1.3$\pm$0.25 & 0.99$\pm$0.14    & 25.2$\pm$14.7 &     37.7$\pm$21.1  & {\it 0.7$\pm$0.2}      \\
Q Cyg    &   0.42 & 113 & 1320$\pm$43 &    3$\pm$0.1 & 14.9$\pm$0.6 &  11$\pm$1 &   5$\pm$1 & 0.81$\pm$0.19 & 0.45$\pm$0.16    &  7.3$\pm$3.19  &    13.1$\pm$5.5 & 1.13$\pm$0.15       \\
HR Del   &   0.214 &  20 &  932$\pm$31 &  3.8$\pm$0.5 & 12.1$\pm$0.1 & 230$\pm$4 & 160$\pm$3 & 0.53$\pm$0.06 &  0.7$\pm$0.07   &  237$\pm$34.5 &   356.9$\pm$50 &  {\it 0.6$\pm$0.1}      \\
DN Gem   &   0.128 &  78 & 1318$\pm$146 &  3.5$\pm$0.1 & 15.5$\pm$0.5 &  37$\pm$3 &  17$\pm$2 & 0.53$\pm$0.12 & 0.61$\pm$0.17  &  5.3$\pm$1.5 &      8.6$\pm$2.4  & 0.93$\pm$0.15     \\
DQ Her   &   0.194 &  52 &  494$\pm$6 & 1.35$\pm$0.1 & 14.3$\pm$0.3 &  94$\pm$6 &  67$\pm$4 & 0.16$\pm$0.06 &  1.5$\pm$0.03    &  2.8$\pm$0.41 &      5.4$\pm$0.7 &  {\it 0.66$\pm$0.1}       \\
V446 Her &   0.207 &  32 & 1308$\pm$130 &    3$\pm$0.1 & 16.9$\pm$0.8 &  15$\pm$2 &   6$\pm$2 & 1.18$\pm$0.12 & 0.99$\pm$0.14  &  2.4$\pm$0.70 &       3.4$\pm$0.9 &  1.09$\pm$0.15    \\
V533 Her &   0.148 &  23 & 1165$\pm$44 &    3$\pm$0.2 & 14.5$\pm$0.5 &  44$\pm$2 &  25$\pm$2 & 0.09$\pm$0.06 & 0.99$\pm$0.1    &  6.8$\pm$0.99  &    12.3$\pm$1.7 &  {\it 0.95$\pm$0.15}       \\
CP Lac   &   0.145 &  55 & 1129$\pm$54 &  2.1$\pm$0.1 & 15.5$\pm$0.3 &  10$\pm$1 &   5$\pm$1 & 0.87$\pm$0.19 & 1.05$\pm$0.09   &  7.2$\pm$3.1  &   11.7$\pm$4.9 &  1.14$\pm$0.15      \\
DI Lac   &  0.544 &  80 & 1569$\pm$51 &  4.6$\pm$0.1 & 14.7$\pm$0.3 &  40$\pm$3 &  20$\pm$2 & 0.81$\pm$0.12 & 0.26$\pm$0.17    & 16.4$\pm$4.8  &   27.4$\pm$7.7 &   0.91$\pm$0.2        \\
DK Lac   &   0.13 &  41 & 2296$\pm$391 &  5.7$\pm$0.3 & 16.8$\pm$0.4 &  60$\pm$15 &  40$\pm$10 & 0.68$\pm$0.19 & 0.72$\pm$0.26 &  2.6$\pm$1.1  &     3.6$\pm$1.5 &   0.83$\pm$0.15      \\
HR Lyr   &    0.1 &  71 & 4493$\pm$684 &  6.5$\pm$0.1 & 15.5$\pm$0.3 &  80$\pm$10 &  45$\pm$7 & 0.56$\pm$0.19 & 0.52$\pm$0.26  &    3$\pm$1.3  &     5.5$\pm$2.3 &  0.78$\pm$0.15      \\
BT Mon   & 0.339 &  51 & 1413$\pm$97 &   & 15.8$\pm$0.6 &   &   & 0.74$\pm$0.19 & 1.45$\pm$0.05      &  2.9$\pm$1.3  &     4.4$\pm$1.8 &  1.05$\pm$0.1          \\
GI Mon   &   0.18 &  72 & 2849$\pm$460 &  5.4$\pm$0.2 & 15.8$\pm$0.8 &  30$\pm$7 &  20$\pm$5 & 0.31$\pm$0.12 & 0.78$\pm$0.26   &  1.9$\pm$0.55  &      3.5$\pm$1.0 &  0.95$\pm$0.1      \\
V841 Oph &  0.601 & 139 &  805$\pm$18 &  4.2$\pm$0.3 & 13.5$\pm$0.3 & 130$\pm$10 &  54$\pm$7 & 1.36$\pm$0.19 & 0.52$\pm$0.17   &  130$\pm$56.7 &  198$\pm$83 &  0.71$\pm$0.15    \\
GK Per   &  1.997 &  85 &  437$\pm$8 &  0.2$\pm$0.1 & 13.4$\pm$0.4 &  13$\pm$1 &   6$\pm$1 & 1.05$\pm$0.12 & 1.24$\pm$0.09     & 11.9$\pm$3.5 &    30.6$\pm$8.9 &  {\it 1.0$\pm$0.2}             \\
RR Pic   &   0.145 &  60 &  504$\pm$8 &    2$\pm$0.3 &   12$\pm$0.2 & 250$\pm$30 & 125$\pm$20 &    0$\pm$0.06 & 1.19$\pm$0.09  & 72.4$\pm$5.3 &    118$\pm$16 & 0.62$\pm$0.2    \\
CP Pup   &    0.061 &  47 &  795$\pm$13 &  0.4$\pm$0.15 &   15$\pm$0.5 &   8$\pm$1 &   5$\pm$1 & 0.62$\pm$0.12 & 0.61$\pm$0.26 &  9.8$\pm$2.8 &    16.9$\pm$4.7 &  1.16$\pm$0.2     \\

\hline
\end{tabular}

\tablefoot{Basic data (columns 2-8 and 10) from: Bode \& Evans (2008), Bruch \&
    Engel (1994), Cohen \& Rosenthal (1983), Collazzi et al (2009), Diaz
    \& Bruch (1997), Downes \& Duerbeck (2000), Downes et al (2001),
    Duerbeck (1981), Duerbeck (1987), Kube et al (2002),
    McLaughlin (1960), Ringwald et al (1996), Ritter \& Kolb (2011),
    Robinson (1975), Shafter (1997), Sparks et al (2000), Strope et
    al (2010), Warner (1985), Warner (1987). Columns 9, 11, and 12 from Paper I,
    column 13 from the present paper. See also Sect. 8 for a discussion of
    individual values in column 10.}

\end{center}
\end{table*}

Table 1 contains the basic data for the novae of our sample obtained
both from the literature (e.g., the magnitudes $m$v$_{min}$ and
$m$v$_{max}$, the rates of decline $t_2$ and $t_3$, the orbital period
$P$, the system inclination, and the time $\Delta T$ elapsed from the
OB at time of the UV observations, etc.) and from Paper I (e.g.,
the reddening $A_{\rm v}$ and the UV-integrated flux
distribution, etc).  The orbital periods are mostly from Ritter \&
Kolb (2011).  The orbital period of HR Lyr is quite uncertain; we
adopted P=0.1$^{\rm d}$ from Leibowitz (1995). For a discussion of the
data regarding the system inclination see Sect. 8.

An important aspect of this paper is the determination of the
uncertainties in the values of the basic physical quantities and their
propagation to the final, most significant parameters, for example  the
accretion luminosity and $\dot{M}$.  All basic and derived
parameters in this study contain error bars. For the basic parameter
data ($m$v$_{min}$, $m$v$_{max}$, $t_3$, P$_{orb}$, etc.), we assumed
as nominal value the average of the various values found in the
literature (with some degree of personal judgement, {e.g.,}
  in identifying multiple values in different old publications coming
  from the same source, sometime without attribution). We considered
as "error bar" the semi-difference of their range; we are aware that
in doing so we are probably overestimating the errors, but using the
range rms as if the values were homogenous and normally distributed
did not appear to be warranted, and we preferred to follow
McLaughlin's (1941) precept of ``erring on the side of conservatism"
when dealing with error estimates.

We found in Paper I that the most important source of error in the
estimate of the UV SED, which is fundamental for the calculation of
the disk luminosity, derives from the uncertainty in the estimate of
the reddening correction, since this quantity directly affects the
value of the index $\alpha$ in the power-law approximation of the SED
and, as a consequence, the $\lambda-$integrated flux.  See Paper I for
the relevant values.

A detailed description of the handling of the propagation of errors is
in Appendix A.

\subsection{A note on the fits in the figures}

There are some figures showing correlations in this paper, and they
include linear fits between variables (or their log). We have chosen
to show both the direct ($y$ vs. $x$) and inverse ($x$ vs. $y$) fits
(dotted lines) because in most instances it is not obvious which
variable is the independent one and we felt that this way gives a
clearer idea of the range of slopes. We also show Deming (1943)
regressions (as dashed lines), which account for errors in both $x$ and $y$
and therefore are, in our opinion, more realistic than standard
$1/\sigma_y^2$ weighted fits.  The fit coefficients in the text refer
to the Deming regressions.

Whenever possible, novae are identified in the figures by unique
three-letter labels.

\subsection{Comments on parameters derived from the WD mass}

Physical parameters that depend on the WD mass are derived in
  this paper for two separate cases: that of an identical white dwarf
  mass ($M_{WD}\equiv1.0 \, M\odot$) for all objects, and that of
  individual white dwarf masses, as determined from the literature and
  from the methods described in Sect. 10.

  For clarity, the parameters are identified with subscripts $1M\odot$
  or $M_{WD}$, respectively.

  In this way we can explore the role played by the WD mass on the
  parameters depending on it, and check for possible bias associated
  to this assumption (in particular avoiding the possible degeneracy
  between the WD mass and the parameters, for example $t_3$, used to derive
  it): if a correlation exists for the 1 M$\odot$ case it is highly
  unlikely that the one for the individual WD masses is due to
  parameter degeneracy.

\section{The distances  in  Gaia  Data Release 2}

The Gaia Data Release 2 (Gaia Collaboration et al. 2018)
provides precise positions, proper motions, and parallaxes for an
unprecedented number of objects, that is,  more than 1.3 billion (Luri et
al. 2018). Data for all 18 novae of our sample are contained in this
release. Their uncertainties are in most cases below 20\%\ and
this would allow a simple inversion of the parallax to derive the
corresponding distance.  However, following the considerations
contained in Luri et al. (2018) and Bailer-Jones et al.  (2018), we avoided this "naive" approach, which is allegedly a biased and
very noisy estimator and tends to overestimate the true
distance. Therefore, the assumed distance values are the point
distance estimates $r_{est}$ in Table 1 of Bailer-Jones (2018),
contained in the table of geometric distance under the schema
gaiadr2-complements at the Gaia TAP service of the Astronomisches
Rechen Institut (ARI), http://gaia.ari.uni-heidelberg.de/tap.html.
The comparison between the 1/$\pi$ and the $r_{est}$ distance
estimates indicates a progressive relative increase in the difference
ratio ($1/\pi -r_{est})/r_{est}$ with larger distances, for example, from
about 1\% at 300 pc to about 10\% near 3000 pc. It is worth
  noting that distances from the nebular expansion parallax for T
Aur, V446 Her, V533 Her, and CP Lac are in fair agreement with Gaia
distances

It is instead surprising, and worth reporting, that the Gaia distances
of V603 Aql ($311\pm7$) and DQ Her ($494\pm6$) are in noticeable
contrast with those derived from the HST-Fine Guidance Sensor,
  both $\sim 25\%$ lower at $249\pm8$ and $386\pm31,$ respectively;
see Harrison et al. (2013).

The  Gaia $r_{est}$ distances from Bailer-Jones et al.  (2018),
with the corresponding errors, are in column 4 of Table 1.

\section{The nova magnitude at maximum }

This quantity is derived from $m$v$_{max}$, $A$v, and the distance
using the common relation $M$v$_{max}$ =
$m$v$_{max} +5 -5 \log d - A$v.  The three quantities on the right-hand
side have uncertainties that are outlined below:

\begin{itemize}
\item{$m$v$_{max}$ :} Due to the uncertain definition of the peak
  visual magnitude. Here, the uncertainties are higher for objects
  with faster decline. For the peak magnitude we used the
  average of the values from the literature.
\item{$A$v :} Here one can find large uncertainties even for well-studied objects, depending on the method adopted: that is, UV bump, IR
  maps, or Balmer decrement.  We used the visual extinction Av
  (Paper I) from a homogeneous set of good-quality data, by the
  method of removal of the wide interstellar dust UV absorption bump
  centered around 2200\AA.
\item{Distance} : The Gaia astrometric distances have errors
  that derive from the uncertainties in the parallax and possible
  systematic errors; see Bailer-Jones et al.  (2018) and Riess et
  al. (2018) for details. In Table 1 we assumed as error the
  semi-difference of $r_{est}^{upper} - r_{est}^{lower}$.
\end{itemize}

The derived values of $M$v$_{max}$ are reported in column 1 of Table
2.  The average of the absolute magnitude at maximum for the novae in
the sample (BT Mon excluded; see Sect. 11.2) is -7.51 $\pm$ 0.96.
This value is in excellent agreement with the results of Shafter et
al. (2011) and Shafter (2013, 2017) who found -7.5$\pm$ 0.8 for the
average magnitude of  Galactic novae.

\begin{table*}
  \caption{Results. For clarity in the table, the errors in the log
    values were computed as $\delta x/x$ even though the condition
    $\,\delta x\ll x\,$ is not always met. }
\begin{center}
\renewcommand{\tabcolsep}{0.16cm}
\renewcommand{\arraystretch}{1.2}
\begin{tabular}{lrrrrrrrrr}
\hline
\smallskip

& \multicolumn{1}{c}{(2)} & \multicolumn{1}{c}{(3)} & \multicolumn{1}{c}{(4)} &
\multicolumn{1}{c}{(5)} & \multicolumn{1}{c}{(6)} & \multicolumn{1}{c}{(7)} & \multicolumn{1}{c}{(8)} &
\multicolumn{1}{c}{(9)} & \multicolumn{1}{c}{(10)} \\
 Object & \multicolumn{1}{c}{\bf $M$v$_{max}$} &
\multicolumn{1}{c}{\bf  $L^{bol}_{max}/L_{\rm Edd}$}
&\multicolumn{1}{c}{\bf $L_{\rm IUE}$} &
 \multicolumn{1}{c}{\bf $L_{\rm disk}^{\rm ref}$}  & \multicolumn{1}{c}{\bf $\dot{\rm M}_{1M\odot}$ } &
\multicolumn{1}{c}{$\log{\dot{\rm M}_{1M\odot}}$} & \multicolumn{1}{c}{\bf $\dot{\rm M}_{M_{WD}}$} &
\multicolumn{1}{c}{$\log{\dot{\rm M}_{M_{WD}}}$ }
& \multicolumn{1}{c}{\bf $M$v$_{min}^{\rm ref}$ }   \\

& \multicolumn{1}{c}{[mag]} & \multicolumn{1}{c}{[$L_{\rm Edd\_1M\odot}$]} & \multicolumn{1}{c}{[$L\odot$]} & \multicolumn{1}{c}{[$L\odot$]} & \multicolumn{1}{c}{$10^{-9} M_{\odot}\,$/yr} & &
\multicolumn{1}{c}{$10^{-9} M_{\odot}\,$/yr} & & \multicolumn{1}{c}{[mag]}\\
\hline

V603 Aql & $-$9.11$\pm$0.21  & 11.0$\pm$2.2 &  5.23$\pm$0.79 &  3.2$\pm$0.5 &  1.69$\pm$0.26 &  $-$8.77$\pm$0.07 &  1.14$\pm$0.52 & $-$8.94$\pm$0.20 &  4.92$\pm$0.12 \\
T Aur    & $-$6.76$\pm$0.28  & 1.26$\pm$0.33 &  5.76$\pm$3.40 &  8.9$\pm$6.9 &  4.68$\pm$3.60 &  $-$8.33$\pm$0.33 & 10.1$\pm$9.3 & $-$8.00$\pm$0.40 &  4.22$\pm$0.52 \\
Q Cyg    & $-$8.41$\pm$0.22  & 5.76$\pm$1.21 &  3.96$\pm$1.75 &  3.2$\pm$1.4 &  1.68$\pm$0.74 &  $-$8.77$\pm$0.19 &  1.21$\pm$0.73 & $-$8.92$\pm$0.26 &  4.30$\pm$0.64 \\
HR Del   & $-$6.58$\pm$0.50  & 1.06$\pm$0.50 & 64.2$\pm$10.2 & 59$\pm$11 & 30.7$\pm$5.9 &  $-$7.51$\pm$0.08 & 81.6$\pm$27.3 & $-$7.09$\pm$0.15 &  2.26$\pm$0.16 \\
DN Gem   & $-$7.63$\pm$0.28  & 2.80$\pm$0.75 &  2.86$\pm$1.04 &  2.5$\pm$1.0 &  1.29$\pm$0.54 &  $-$8.89$\pm$0.18 &  1.52$\pm$0.84 & $-$8.82$\pm$0.24 &  5.02$\pm$0.60 \\
DQ Her   & $-$7.28$\pm$0.12  & 2.02$\pm$0.24 &  0.21$\pm$0.03 &  6.5$\pm$3.9 &  3.43$\pm$2.05 &  $-$8.46$\pm$0.26 &  7.77$\pm$5.06 & $-$8.11$\pm$0.28 &  2.89$\pm$0.66 \\
V446 Her & $-$8.76$\pm$0.26  & 7.96$\pm$2.00 &  1.28$\pm$0.45 &  1.9$\pm$1.0 &  1.00$\pm$0.51 &  $-$9.00$\pm$0.22 &  0.80$\pm$0.51 & $-$9.10$\pm$0.28 &  5.12$\pm$0.90 \\
V533 Her & $-$7.42$\pm$0.22  & 2.31$\pm$0.50 &  2.87$\pm$0.47 &  5.2$\pm$1.5 &  2.72$\pm$0.80 &  $-$8.57$\pm$0.13 &  3.07$\pm$1.41 & $-$8.51$\pm$0.20 &  4.07$\pm$0.57 \\
CP Lac   & $-$9.03$\pm$0.23  & 10.2$\pm$2.3 &  2.85$\pm$1.27 &  5.7$\pm$3.0 &  2.97$\pm$1.58 &  $-$8.53$\pm$0.23 &  2.07$\pm$1.41 & $-$8.68$\pm$0.30 &  4.22$\pm$0.43 \\
DI Lac   & $-$7.19$\pm$0.17  & 1.86$\pm$0.31 & 12.5$\pm$3.8 &  7.8$\pm$2.5 &  4.09$\pm$1.30 &  $-$8.39$\pm$0.14 &  5.07$\pm$2.87 & $-$8.29$\pm$0.25 &  3.84$\pm$0.34 \\
DK Lac   & $-$6.78$\pm$0.51  & 1.29$\pm$0.60 &  4.27$\pm$2.37 &  3.7$\pm$2.6 &  1.92$\pm$1.36 &  $-$8.72$\pm$0.31 &  2.86$\pm$2.27 & $-$8.54$\pm$0.34 &  4.83$\pm$0.68 \\
HR Lyr   & $-$7.32$\pm$0.39  & 2.11$\pm$0.77 & 18.9$\pm$10.0 & 16.9$\pm$10 &  8.82$\pm$5.22 &  $-$8.05$\pm$0.26 & 14.8$\pm$10.3 & $-$7.83$\pm$0.30 &  2.42$\pm$0.54 \\
BT Mon   &                 &               &  1.80$\pm$0.82 & 24$\pm$17.6 & 12.6$\pm$9.2 &  $-$7.90$\pm$0.32 & 11.1$\pm$8.6 & $-$7.95$\pm$0.34 &  2.20$\pm$0.83 \\
GI Mon   & $-$7.18$\pm$0.42  & 1.85$\pm$0.73 &  4.80$\pm$2.08 &  6.0$\pm$3.7 &  3.12$\pm$1.96 &  $-$8.51$\pm$0.27 &  3.52$\pm$2.36 & $-$8.45$\pm$0.29 &  3.62$\pm$0.98 \\
V841 Oph & $-$6.69$\pm$0.35  & 1.18$\pm$0.40 & 26.2$\pm$11.5 & 18.8$\pm$9.7 &  9.85$\pm$5.90 &  $-$8.01$\pm$0.26 & 19.6$\pm$12.5 & $-$7.71$\pm$0.28 &  3.36$\pm$0.39 \\
GK Per   & $-$9.05$\pm$0.16  &10.4$\pm$1.6 &  0.70$\pm$0.20 &  4.1$\pm$1.9 &  2.14$\pm$0.99 &  $-$8.67$\pm$0.20 &  2.14$\pm$1.43 & $-$8.67$\pm$0.29 &  3.36$\pm$0.55 \\
RR Pic   & $-$6.51$\pm$0.30  & 1.00$\pm$0.29 &  5.73$\pm$0.45 & 18.1$\pm$6.7 &  9.44$\pm$3.51 &  $-$8.02$\pm$0.16 & 23.7$\pm$15.5 & $-$7.62$\pm$0.28 &  2.91$\pm$0.38 \\
CP Pup   & $-$9.72$\pm$0.19  &19.2$\pm$3.6 &  1.92$\pm$0.56 &  1.8$\pm$0.8 &  0.92$\pm$0.40 &  $-$9.03$\pm$0.19 &  0.61$\pm$0.44 & $-$9.22$\pm$0.32 &  5.53$\pm$0.60 \\

\hline
\end{tabular}

\end{center}
\end{table*}

\section{ The MMRD  revisited with the new Gaia distances}

The empirical relation between the absolute optical magnitude at
maximum and the rate of decline $t_n$ of a nova light-curve (MMRD)
provides insight into the nova phenomenon (faster novae are
intrinsically brighter than slower ones) and is a convenient method
for estimating the distance of Galactic novae in the absence of
other more direct estimates (although possibly not as relevant in the
Gaia era). The MMRD gives an estimate of $M$v$_{max}$, and hence the
distance can be derived if the apparent magnitude at maximum, the
$t_n$ values, and the visual extinction are known.  It should be
  stressed, however, that with Gaia the MMRD becomes better calibrated
  than ever before, making it an invaluable tool for extragalactic
  studies.

Since the pioneering work by McLaughlin (1945) a number of MMRD
relations for Galactic and extragalactic novae have been
proposed by different authors (see della Valle \& Gilmozzi 2002 for a
review), the most recent ones being those of the S-shaped curve of
della Valle \& Livio (1995), and the linear relation of Downes \&
Duerbeck (2000). The first theoretical explanation of the MMRD
  relationship for CNe was proposed by Shara (1981).  More recently,
Hachisu \& Kato (2006, 2007, 2010) proposed a theoretical explanation
of the general trend observed in the empirical MMRD curve, in the
context of the "universal decline law" Mv = 2.5 log $t_3$ -11.6.

The validity of the MMRD relation for extragalactic novae, where the
uncertainties in the distances are much lower, was confirmed by
Darnley et al. (2006),  Shafter et al, (2011), Shafter et
al. (2012), and Shafter (2013) from a photometric and spectroscopic
study of a great number of novae in M31, M33, and the  Large
Magellanic Cloud. The peak nova
luminosity appears clearly correlated with the rate of decline,
that is, the more luminous novae generally fade the fastest.

Instead, Kasliwal et al. (2011), in a study of a more limited number
of extragalactic novae found  a subclass of faint and fast objects
which fall  below the MMRD relation.  The presence of
faint and fast
  extragalactic novae was confirmed by the observation of the
  recurrent nova M31-12a (Darnley et al. 2015) and by the survey of
  novae in M87 by Shara et al. (2017). Also, Cao et al. (2012), in a
study of PTF- and Galex-based light curves of 29 novae in M31, a subset
of all the CNe in M31, found significant scatter and a number of
outliers in the MMRD distribution.

While some of this scatter is possibly due to the uncertainties
in the determination of the visual extinction, in the case of
  Galactic novae there may be substantial uncertainties in the
determination of distances from expansion parallaxes when the ejecta
are not spherically symmetric (see Wade et al. 2000 for general
considerations on this topic).  Shore (2012, 2013, 2014) also pointed
out the role played by the aspherical geometry of the ejecta near
maximum light, with a range of opening angles and inclinations, on the
observed scatter in the MMRD relation.

In view of the above, it is understandable that the availability of
the new Gaia distances represents an irresistible
temptation to test and recalibrate the MMRD relation using the data of
our 17 novae with the reliable Gaia distances and the
homogeneous E(B-V) values.  We excluded BT Mon from this sample
because of the uncertainties in its rate of decline, $t_3$ , and its
$m$v$_{max}$ value (see Sect. 11.2).

The data in the MMRD diagram are affected by the uncertainties in
$M$v$_{max}$ discussed in Sect. 4, and by those in $t_3$, that is, the
uncertainty on the exact time and value of the maximum light and,
because of its ambiguous definition, by the uncertainty in
estimating the magnitude at $t_3$ due to the frequent presence of
jitter and other fluctuations in the light curve, whose decline is
rarely represented by a smooth curve (Strope et al. 2010; Burlak \&
Henden 2008).

The $t_3$ values in column 7 of Table 1 were derived from the
literature as our best estimates. As mentioned above, both the
determination of the peak luminosity and that of $t_3$ is not a
straightforward procedure, especially for objects with jitter in the
light curve (see Burlak and Henden, 2008 and Strope et al. 2010 for
details).  Burlak and Henden (2008) derived the photometric parameter
of their sample of novae from smoothed light curves that ignore small
flares.  Instead, Strope et al. (2010) in their classification of nova
light curves choose to use the {observed} light curve and to use
for $t_3$ the last time in the light curve for which the brightness
was above the threshold. We also note that Strope et al.  (2010)
consider the jitter as "extra light" added onto the decline law.

In our sample of stars, uncertainties associated with the presence of
jitter are especially evident in RR Pic. In this case, we followed the
method of Burlak and Henden (2008) and the procedure of Darnley et
al. (2006) for novae in M31 and derived the $t_3$ data from
interpolation between points on the decline of the light curve, that
is, from a smoothed curve between spikes, a method similar to the
tracing of the continuum in a noisy spectrum.  Careful examination
of the first stages of the OB light curve of RR Pic, from the
data published by Spencer Jones (1931) and Lund (1926) led to
an estimate of the maximum magnitude of close to 2.0, and to $t_2$ and $t_3$
values close to 125 and 250 days, respectively.

\begin{figure}
\centering
\resizebox{\hsize}{!}{\includegraphics   [angle=0] {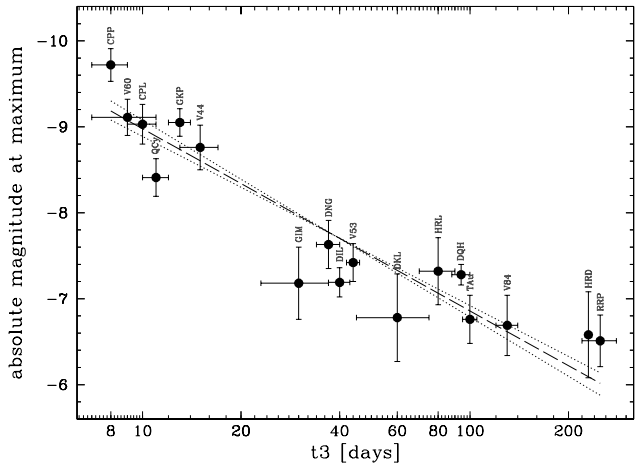}}
\caption{A revision of the   MMRD relation,  using
the new distances from Gaia DR2. See Sect. 2.1 for explanation of the fit lines.}
\end{figure}

DK Lac is another object with an OB light curve characterized by
jitter. The commonly reported value of $t_3$ in the literature varies
between 24 and 32 days (Duerbeck 1981; Cohen 1985; Duerbeck \& Seitter
1987; Downes \& Duerbeck 2000). However, inspection of the original
studies (see Ribbe (1951), Wellmann (1951) Bertaud \& Baldet (1952),
Larsson-Leander (1954)) indicates that these $t_3$ values derive from
an arbitrary extrapolation of the peak magnitude to a value about 0.5
mag brighter. Instead, Strope et al.  (2010), based on AAVSO
observations, give an exceptionally long $t_3$ value of 202 days. This
probably derives from their treatment of the presence of jitter.  In
our estimate, by adopting a method similar to that described above, we
derived a $t_3$ value of 60$\pm$15 days, intermediate between the two
extremes.

For the recalibration of the MMRD relation we adopted the
$M$v-$\log t_3$ "universal" decline law relation.  The new MMRD
(Fig. 2) is
\begin{equation}
M{\rm v}_{max} = (2.12\pm 0.20) \cdot \log t_3 - 11.08\pm 0.33
.\end{equation}

\noindent Even with only 17 data points this MMRD has a tight
correlation ($r=0.93$, rms = 0.38 mag) and thanks to the superior
precision of the Gaia distances, confirms that the linear fit
between $\log t_3$ and $M$v$_{max}$ is the appropriate functional
dependence.  We believe that this result strengthens the validity of
the universal decline law as the correct description of the MMRD.

While the current paper was very close to submission, a study on nova
distances and MMRD appeared (Schaefer 2018) in which the author
concludes that the MMRD relation is too poor and should no longer be
used.

We note here that out of our sample of 18 objects, 15 are in
common with the larger sample of Schaefer (2018) and that for 10 of
these objects our parameter values are close to those of Schaefer. For
V603 Aql and V446 Her we have different $m$v$_{max}$, while for V446
Her, DK Lac, and RR Pic we have different $t_3$ values. As explained
above, our $m$v$_{max}$ and $t_3$ values come from careful
examination of many sources in the literature data (including Strope,
Schaefer, and Henden 2010) and all contain error bars, unlike those in
Schaefer (2018). The difference in the $t_3$ values may perhaps be
explained by the different method to estimate $t_3$ when jitter is
present (as in the case e.g., of DK Lac and RR Pic).

Unlike Schaefer (2018) we have {not} included BT Mon in
our MMRD because of the uncertainty in its $m$v$_{max}$ and $t_3$
values (see Sect. 11.2 for a more detailed discussion). Schaefer
instead included BT Mon in his ``gold'' sample, declared it a
``confident outlier'' and used it as a particularly clear example
against the MMRD.

\section{The absolute magnitude 15 days after maximum}

Buscombe \& de Vaucouleurs (1955) noted that about 15 days after
maximum all novae have an absolute magnitude ($M$v$_{15}$) close to
$-5.2 \pm 0.1$, independent of the speed class. More recent studies
are those of Pfau (1976) ($M$b$_{15}= -5.74 \pm 0.60$), Cohen (1985)
($M$v$_{15} = -5.60 \pm 0.45$), van den Bergh \& Younger (1987)
($M$v$_{15} = -5.23 \pm 0.16$), Capaccioli et al. (1989)
($M$v$_{15} =-5.69 \pm 0.42$) and Downes \& Duerbeck (2000)
($M$v$_{15} =-6.05 \pm 0.44$).

Ferrarese et al. (2003), using extragalactic novae in M49, derived
$M$v$_{15} = -6.36 \pm 0.43$.  Darnley et al. (2006), using the
POINT-AGAPE microlensing survey of M31, derived
$M$r$_{15} = -6.3 \pm 0.9 $ and $M$i$_{15} = -6.3 \pm 1.0$, for the
SLOAN $r^\prime$ and $i^\prime$ filters, but because of the large
scatter concluded that there was no evidence of a $M_{15}$
relationship. Very recently, Shara et al. (2018), from a HST survey for
novae in M87, found instead support for the $M_{15}$ relation and
derived $M$v$_{15}=-6.37 \pm 0.46$.

Following a welcome suggestion by the referee we recalibrated the
$M$v$_{15}$ relation using 17 Galactic novae of our sample (BT Mon
excluded) with Gaia DR2 distances, homogeneous reddening correction
from our UV-based study (PaperI), and $m$v$_{15}$ data from the
literature. For 11 objects the $m$v$_{15}$ values are from Cohen
(1985), Downes \& Duerbeck (2000), and the compilation by Kantharia
(2017). For the remaining 6 objects (Q Cyg, DN Gem, DI Lac, HR Lyr, GI
Mon, V841 Oph) the $m$v$_{15}$ data are derived from the light curves
by Shapley (1933), Cecchini \& Gratton (1942), Payne-Gaposchkin
(1957), and McLaughlin (1960).  We also used the AAVSO light
curves, and additional information from Duerbeck (1987) and Shara et
al. (1989). For the choice of the final values and the estimate of the
errors, for which no information is given in the literature, we have
taken into consideration the uncertainties in $m$v$_{max}$ and its
timing, and the scatter in the light curves, to derive conservative
estimates of the errors.

Column 2 of Table 3 gives $m$v$_{15}$ with associated uncertainties
(typically 0.2-0.3 mag). The $M$v$_{15}$ values with their final
errors (propagated from the relevant data in Table 1) are in column
3. The arithmetic mean is $M$v$_{15} = -5.58 \pm 0.41$, while the
weighted one is $M$v$_{15} = -5.71 \pm 0.40$. A comparison with
earlier data should be made using the first value since all previous
averages were arithmetic ones.

\smallskip \noindent We note that:

\smallskip \noindent 1. The small error of the mean ($\sim 0.40$ mag)
lends support to the use of $M$v$_{15}$ as a standard candle for
Galactic novae.

\smallskip \noindent 2.  Our average values are quite close to those
of previous $M$v$_{15}$ estimates based on Galactic novae, in spite of
the less reliable values for distance and $A_{\rm v}$ used in those
studies.

\smallskip \noindent 3. Our average is substantially different from
that of Shara et al (2018) for novae in M87 ($-6.37 \pm 0.46$) and of
Ferrarese et al (2003) for novae in M49 ($-6.36 \pm 0.43$). In fact,
none of our objects have an individual value as high as the average of
either study.  This may indicate possible issues with the distances to
the two galaxies, or suggest some systematic differences between the
properties of extragalactic CNe and those of our galaxy.

\smallskip \noindent 4. Our average value is close to the theoretical
values derived by Hachisu \& Kato (2015) for 0.7--1.05 $M\odot$ WDs
using their ``CO nova 4'' models ($M$v$_{15} = -5.4 \pm 0.4$) and for
0.7--1.3 $M\odot$ WDs with the ``Ne nova 2'' ones
($M$v$_{15}= -5.6 \pm 0.3$).

\begin{table}
\caption{Apparent and absolute magnitude 15 days after maximum}
\begin{center}
\renewcommand{\tabcolsep}{0.4cm}
\begin{tabular}{lrrrr}

\hline
\hline
\multicolumn{1}{l}{Object} &
\multicolumn{1}{c}{{$m$v$_{15}$}} &
\multicolumn{1}{c}{{$M$v$_{15}$}} \\
\hline
\hline

V603 Aql  &  2.7 $\pm$ 0.2 & $-$5.01 $\pm$ 0.22  \\
T Aur     &  4.8 $\pm$ 0.3 & $-$6.17 $\pm$ 0.40  \\
Q Cyg     &  6.0 $\pm$ 0.4 & $-$5.41 $\pm$ 0.45  \\
HR Del    &  5.1 $\pm$ 0.2 & $-$5.28 $\pm$ 0.22  \\
DN Gem    &  5.8 $\pm$ 0.4 & $-$5.33 $\pm$ 0.48  \\
DQ Her    &  2.5 $\pm$ 0.1 & $-$6.13 $\pm$ 0.12  \\
V446 Her  &  6.0 $\pm$ 0.2 & $-$5.76 $\pm$ 0.32  \\
V533 Her  &  4.3 $\pm$ 0.2 & $-$6.12 $\pm$ 0.22  \\
CP Lac    &  5.6 $\pm$ 0.2 & $-$5.53 $\pm$ 0.29  \\
DI Lac    &  6.0 $\pm$ 0.2 & $-$5.79 $\pm$ 0.25  \\
DK Lac    &  7.6 $\pm$ 0.3 & $-$4.89 $\pm$ 0.51  \\
HR Lyr    &  8.0 $\pm$ 0.3 & $-$5.82 $\pm$ 0.48  \\
GI Mon    &  7.7 $\pm$ 0.3 & $-$4.88 $\pm$ 0.48  \\
V841 Oph  &  5.0 $\pm$ 0.3 & $-$5.89 $\pm$ 0.36  \\
GK Per    &  3.3 $\pm$ 0.2 & $-$5.95 $\pm$ 0.24  \\
RR Pic    &  3.1 $\pm$ 0.2 & $-$5.41 $\pm$ 0.21  \\
CP Pup    &  4.7 $\pm$ 0.2 & $-$5.42 $\pm$ 0.24  \\
\hline
\hline
\end{tabular}
\tablefoot{For the origin of the data, see text in Sect. 6.
Absolute magnitudes
derived using data of Table 1.}
\end{center}
\end{table}

\section{The luminosity at maximum  and the Eddington limit}

It is generally assumed that the maximum emitted luminosity by a
self-gravitating object in hydrostatic equilibrium cannot exceed
$L_{\rm Edd}$.  However, nature has somehow found a way to circumvent
this restriction and novae are well studied systems exhibiting
super-Eddington luminosities for a relatively long period close to
maximum light. See Hayes et al. (1990), Shaviv (1998), and Kato \&
Hachisu (2007) for general considerations.

\begin{figure}
\centering
\resizebox{\hsize}{!}{\includegraphics   [angle=0] {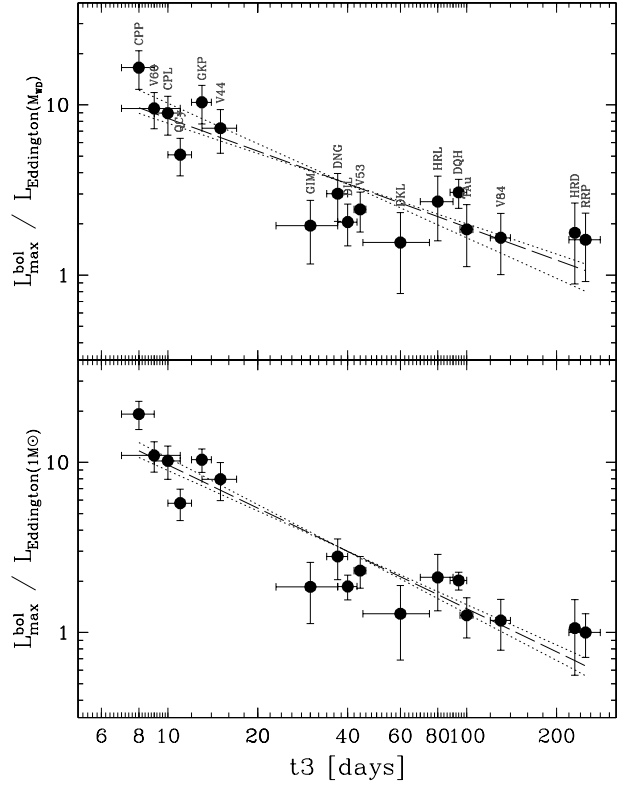}}
\caption{Correlation between the bolometric luminosity at maximum
    (in units of Eddington luminosity) and the speed class $t_3$. {\it
      Top}: $L_{Edd}$ computed using individual WD masses. {\it
      Bottom}: $L_{Edd}$ computed using $M_{WD}=1M\odot$ for each
    nova.}
\end{figure}

The bolometric luminosity at maximum can be derived from
$(M$v$_{max}+BC)$, where we use a bolometric correction of
  $-0.15\pm0.05$ which is intermediate between the values commonly
found in the literature ($-0.25$, Duerbeck 1981; $-0.1$, Livio 1994;
$-0.10$, Duerbeck 1992). Subsequently, $L^{bol}_{max}$ is computed using the
standard relation $L_{\lambda}=3.03\cdot 10^{(35-0.4
  M_{\lambda})}$. We note that the caveats in Sect. 5 on the sources
  of errors and uncertainties in the determination of $M$v$_{max}$ and
  $t_3$ apply here as well.

The value of $L_{Edd}$ can be estimated using the relations given by MacDonald
(1983), $L_{Edd}=2.5\,10^{38} \, (1+X)^{-1}\,M_{WD}$ [erg/s], or by
Warner (1995) and Shaviv (1998),
$L_{Edd} = 6.5 \, 10^4 \, M_{WD} \, (1+X)^{-1}\, $ L$\odot$ [erg/s],
where $M_{WD}$ is in units of M$\odot$ and X is the hydrogen mass
fraction.  For X=0.7 and $M_{WD}$=1M$\odot$,
$L_{Edd} \sim 1.4 \, 10^{38}$ erg/s.  Iben \& Tutukov (1984) give a
slightly different relation:
$L_{Edd} = 4.6\,10^4 (M_{WD}-0.26) \, L\odot$ that gives
$L_{Edd} \sim 1.31\, 10^{38}$ erg/s for a 1.0 M$\odot$ white dwarf.

In the following we adopt $1.4\cdot10^{38}$ erg/s for the average
Eddington luminosity of a 1.0 M$\odot$ WD.  Using the relation by Warner
  (1995) and Shaviv (1998),  $L_{Edd}$ was also
  computed for individual WD masses. Column 3 of Table 2 gives
$L^{bol}_{max}/L_{Edd}$ ratios for $M_{WD}$= 1M$\odot$.

All objects appear super Eddington, with a clear dependence of the
ratio on the speed class, fast objects being strongly super-Eddington;
see also Yaron et al. (2005). For $L_{Edd}(1M\odot)$ the ratio
  varies from $\sim$ 19.2 for the fastest objects, for example CP Pup ($t_3$
$\sim$ 8 days), to $\sim $ 1.0 for the slowest novae, for example  HR Del
($t_3$ $\sim$ 230 days).  The corresponding ratios for
  $L_{Edd}(M_{WD})$ are 16.6 and 1.1.

\smallskip
A clear correlation exists between $L^{bol}_{max}/L_{Edd}$ and
$\log t_3$ (see Fig. 2) whose fits (computed as described in
  Sect. 2.2) give:
\begin{equation}
\log(L^{bol}_{max}/L_{Edd})_{1M\odot}=-0.84\pm0.08\,\,\log t_3 +1.83\pm0.14
,\end{equation}
\begin{equation}
\log(L^{bol}_{max}/L_{Edd})_{M_{WD}}=-0.64\pm0.09\,\,\log t_3 +1.56\pm0.15
.\end{equation}

The fact that such a definite relation exists could appear as trivial,
because $L^{bol}_{max}$ is tightly associated to $M$v$_{max}$,
$L_{Edd}$ is proportional to $M_{WD}$, i.e., almost a constant
($\sim$0.9$\pm$0.2), and the MMRD is a relation between
$M$v$_{max}$ and $t_3$. However, while the MMRD ``describes'' the
fact that faster novae are brighter in a way common to all,
  $L^{bol}_{max}$/$L_{Edd}$  vs. $t_3$ (thanks to the new Gaia distances)
also gives
  physical information: it confirms that all novae are
  super-Eddington, the more so the faster they are.

\section{The   absolute magnitude  at  minimum}

The absolute magnitude at minimum is derived from the apparent
magnitude $mv_{min}$ using the common relation
$Mv_{min} = mv_{min} +5 - 5 \log d -A_{\rm v}$.

Warner (1986, 1987) defined this $Mv_{min}$ as the apparent
absolute magnitude since in CVs the observed flux depends on the
system inclination angle.

Assuming that the visual radiation originates in a nonirradiated disk
(see the results of Paper I), the "apparent" absolute magnitude
$Mv^{obs}$ must be corrected for the inclination angle of the disk
(see Warner 1987) by a term $\Delta Mv(i)$ to obtain an absolute
magnitude averaged over all directions, variously defined as the
"standard", "reference", or "average" absolute magnitude (we
use "reference" in the following):
\begin{equation} Mv^{ref}=Mv^{obs}-\Delta Mv(i)
,\end{equation}
\noindent where $\Delta Mv(i)$=$-2.5\log[(1+1.5\cos i)\cos i]$
according to Warner (1987).  This relation is derived from a more
general $\lambda$-dependent prescription of Paczynski \&
Schvarzenberg-Czerny (1980) in the specific case of a limb-darkening
parameter $f^{-1}\sim0.6$ for the V region (see Sect. 9).  The
$\Delta Mv(i)$ correction accounts for both geometrical and
limb-darkening effects.

The visual reference magnitude would correspond to the "apparent" one
if viewed at an angle of about 57 degrees.

The correction for inclination to be applied to the apparent visual
magnitude can reach $-5$ magnitudes for systems seen at high
inclination angles (eclipsing objects) while that for systems seen
nearly pole-on is of about +1 magnitude.

Regrettably, the system inclination angle is, in general, not
accurately determined, except for the few systems that exhibit
definite or grazing eclipses.  It is a fortunate circumstance,
however, that the correction for inclination is critical only for
systems with high inclination (eclipses) while it is not so critical
for systems with uncertain (i.e., mid or low) inclination.

The adopted values for the inclination (see Table 1) are mostly from
the compilation by Ritter \& Kolb (2011), complemented with other
information from the literature, i.e., Warner (1987), Duerbeck (1992),
Downes \& Duerbeck (2000), Peters \& Thorstensen (2006), Puebla et al
(2007), and Darnley et al. (2012).  We assumed our
best estimate from the various values in the literature as the nominal value of $i$ , with an
indicative "error" given by half the range of values. The errors
extend to $\pm15$ degrees for systems with uncertain inclination.
A typical magnitude range associated with this error depends on
  the inclination, and varies between 0.5 (i=30) and 0.9 mag (i=45).

Column 10 of Table 2 gives the reference absolute magnitude
$M$v$^{ref}_{min}$ whose average value is 3.88 $\pm$ 1.01.  The
intrinsically most luminous object in quiescence is BT Mon.

\paragraph{\sl Comments on the inclination of individual objects\\}
\verb| |

\noindent V446 Her, CP Lac, Q Cyg, DK Lac, HR Lyr and GI Mon lack reliable
inclination values from the literature.  However, from the absence of
eclipses one can set an upper limit of about 60 degrees for $i$.  In a
few cases, in order to derive an estimate of $i$ we employed the
method of Warner (1986), who pointed out that a definite correlation
exists between the system inclination angle and the equivalent width
of the hydrogen and helium emission lines observed in
  quiescence.
Comments on individual objects are below.

\begin{enumerate}

\item In the case of V446 Her, an additional constraint comes from the
  high K1 value (about 106 km/s) reported by Thorstensen and Taylor
  (2000).  Inserting this value in the mass function, with P=0.207,
  $M_{WD}$=1.09, $M_2$=0.46 M$\odot$ (from the Period-$M_2$ relation,
  see Knigge 2006; Selvelli \& Gilmozzi 2008) one derives that $i$
  $\sim $57 ($i =55$ if $M_{WD}$=1.0).  We adopted 57$\pm$8 degrees.

\item For CP Lac, similar considerations are based on the K1 value
  (about 100 km s$^{-1}$) reported by Peters \& Thorstensen (2006).
  Inserting this K1 in the mass function, with P=0.145, $M_{WD}$
  $\sim$ 1.0, $M_2$=0.26 M$\odot$ one derives $i$ values in the range
  55-65 degrees, with a limited range for reasonable $M_{WD}$
  values. Smaller values of $i$ would require unlikely small values of
  K1, less than 80 km/s, not compatible with the observed RV
  amplitude.  We therefore adopted $i = 60\pm$5 degrees. This agrees
  with the possible detection of shallow eclipses in the light curve
  from time-resolved photometry of the system (Rodriguez-Gil \&
  Torres, 2005).

\item For Q Cyg, Warner (1987) gives $i$ $\sim$ 50, while Peters \&
  Thorstensen (2006) give $i \sim$ 26$\pm$9. The optical spectrum of
  Ringwald et al. (1996) shows weak H-alpha emission. We adopted
  26$\pm$9 degrees.

\item For DK Lac we adopted 41$\pm$15 from Takei et al. (2013) and
  Honeycutt et al. (2011).

\item For HR Lyr we adopted 30$\pm$15 based on the Warner's (1987)
  estimate (0-30) and on the presence, although not prominent, of H
  and helium emission lines in the spectrum (Harrison et al. 2013).

\item For GI Mon we adopted 45$\pm$15.  The optical spectrum (see
  Bianchini et al. 1991, 1992, and Liu \& Hu (2000)) shows weak
  hydrogen and HeII 4686 emission lines.  A recent SALT spectrum of GI
  Mon (Tomov et al. 2015) also shows weak HeII 4686 and H-beta
  emission lines. The equivalent width of H-alpha suggests a
  moderately low inclination.  This seems confirmed by the presence of
  modulations in the light curve ( Woudt et al.  2004) .

\item The inclination of V841 Oph is also uncertain: Diaz \& Ribeiro
  (2003) give a wide possible range (38$\pm$30) while Peters \&
  Thorstensen (2006) give 30$\pm$10;  we adopted the latter value.

\end{enumerate}

\section {The  UV and optical disk    luminosity}

One of the main results of Paper I was the observational
  confirmation that the UV flux, whose SED is well described by a
single power-law (PL) distribution, comes from an accretion disk that
accounts also for the observed optical flux.

Therefore, one can estimate the total disk flux, or at least provide a
reliable (lower) limit for the sum of the UV and optical contributions
by integrating the power law from 1100 to 6000\,\AA.

We reiterate, from Paper I, that the PL indexes were derived using the
Cardelli et al. (1991) reddening correction and that the difference
between the actual UV flux distribution and the corresponding PL
approximation is small. We also found that the observed optical
magnitude agrees with the extrapolation of the PL to the V band,
although on average $m$v$_{obs}$ is $\sim0.17$ mag fainter than
$m$v$_{PL}$, which we interpreted as due to the fact that the V light
comes from the external region where the disk starts to be optically
thin or has a physical edge. Alternatively it could be due to the
  presence of a Balmer jump, although one is not visible in the
  1000-9000 \AA\ spectrum of V603 Aql; see Fig. 15 of Paper I.

We therefore used the method described above to estimate the
total flux of the disk and to derive its luminosity using the
  Gaia distances of Table 1.

We neglected the IR contribution to the disk luminosity by
truncating the integral of the PL at 6000\AA.  This is justified by
the considerations above.  In any case the PL contribution to this
range would be small.

\paragraph{\sl The  UV correction  for the disk inclination \\}
\verb| |

\noindent If the observed SED (or its PL approximation) derives from a disk, a
correction for inclination similar to that introduced in Sect. 8
for the visual magnitude is required. As mentioned there, Paczynski
\& Schvarzenberg-Czerny (1980) derived a general relation for the
correction of the observed disk luminosity for $i$-related limb
darkening and geometrical factors. The reference disk luminosity is
\begin{equation} L^{disk}_{ref} = L^{disk}_{obs} \,
f_{\lambda}^{-1}(i)
,\end{equation}
where
\begin{equation}
f_{\lambda}^{-1}(i)=(((1-u_ {\lambda}+u_ {\lambda} \,
\, \cos(i))/(0.5-u_{\lambda}/6))^{-1}.
\end{equation}
\noindent Here u$_{\lambda }$ is the limb darkening coefficient, whose
values vary from 0.2 (IR), 0.6 (V), and 0.8 (UV), up to 1.0 (far UV).
Limb darkening is an important effect, especially for flat objects
such as accretion disks, and is especially important in the
UV; see Diaz et al. (1996) for relevant considerations.

For physical reasons and for the sake of homogeneity in the treatment
of the error propagation, rather than using separately the approximate
coefficient in the relevant wavelength range, we chose to use the
general equation of Paczinsky \& Schvarzenberg-Czerny (1980) with a
limb-darkening coefficient that linearly varies with wavelength:
\begin{equation} u(\lambda) = 0.85 -4.1667 \cdot 10^{-5} \,
\lambda,
\end{equation}
\noindent where the two constants were derived by imposing u(1200) =
0.8 and u(5500) =0.6.  Equation (4) can also be re-written as :
\begin{equation} L^{disk}_{ref}/L\odot = 3.116 \cdot 10^4 \, d^2
 \int_{1100}^{6000} F_{\lambda} \, f^{-1}_{\lambda}(i)  \,\,
d\lambda
,\end{equation}
\noindent where $d$ is from Table 1, $f^{-1}_{\lambda} (i)$ is given
by Eq. 5, $F_{\lambda} = A \, \lambda^{-\alpha}$, and the constants
$A$ and $\alpha$ are determined from the PL approximation to the
observed UV flux, as described in Paper I.

Columns 4 and 5 of Table 2 give, separately, the UV $\lambda\lambda$
1250-3100 luminosity, as derived from the IUE reddening corrected
$\lambda-$integrated flux, and the total $i$-corrected UV+optical disk
luminosity as derived from Eq. 7.

\section{The  white dwarf  mass and radius }

A basic paradigm in the theoretical interpretation of the nova
phenomenon (Starrfield et al. 1985; Livio \& Truran 1986; Shara 1989,
Truran \& Livio 1989; Livio 1994; Bode \& Evans 2008) is that the
OB  characteristics, that is, the luminosity at
maximum, the
decline of the nova light curve (speed class), the mass of the ejecta,
the outflow velocities, the OB recurrence time, and so on, depend primarily
on the WD mass, and more weakly on other physical parameters like $\dot{M}$,
the chemical composition of the accreted
material, the temperature of the WD, magnetic fields, mixing processes
in the WD, and so on (see also Yaron et al. 2005; Townsley \& Bildsten 2005;
Kato \& Hachisu 2007; Kato et al. 2013).

The white dwarf mass $M_{WD}$ (with its associated radius $R_{WD}$)
plays an important role also in quiescence because the ratio
$R_{WD}$/$M_{WD}$ explicitly appears in the relation between $\dot{M}$
 and the accretion luminosity; see the following
section.

Regrettably, $M_{WD}$ is not accurately known by direct observations
because of the several and severe problems one encounters in
determining the primary mass and other system parameters from the
observed radial velocity curves. One example is that the velocities may
not be those of the star(s), for example if they originate in or above
the disk (see Wade \& Horne 1988; Thorstensen et al. 1991; Marsh \&
Duck 1996).

Due to the fact that reliable determinations of $M_{WD}$ are
only available for a few of our novae, and that even a rough estimate of $M_{WD}$ is
helpful to more accurately determine the nova OB and quiescent
characteristics, we made an effort to derive the masses of the
remaining objects using both a semi-empirical and an empirical
approach.

Livio (1992b) derived a theoretical relation between $t_3$
and $M_{WD}$ (his Eq. 12: $t_3=A \, X^{-1} \, (X^{-2/3}-X^{2/3})^{3/2}$,
where X=$M_{WD}$/1.433). The relation was calibrated using only the WD mass of
V1500 Cyg, i.e., $1.25 M\odot$, to derive A=51.3.

Using data from Ritter \& Kolb (2011) and additional literature (Horne et
al.  1993; Arenas et al. 2000; Rodriguez-Gil \& Martinez-Pais 2002;
Smith \& Vande Putte 2006; Hachisu \& Kato 2007) on $M_{WD}$ for the
seven calibrators (V603 Aql, T Aur, HR Del, DQ Her, V533 Her, GK Per,
V1500 Cyg) with the "best", though admittedly not completely satisfactory,
determinations of the primary mass, we assumed the shape of the
relation to be correct and recalibrated the constant A in Livio's
equation. The results are only partly satisfactory because the derived
value of the constant A is relatively uncertain: 70$\pm$20.  However, this
range in A corresponds to uncertainties of about $\pm$ 0.1 M${\odot}$
for the WD mass, a result that we consider acceptable.

In a different approach, we fitted the data of the seven calibrators
with a simple linear regression of the form $M_{WD}(t) =a+b \, \log t$
for both $t_2$ and $t_3$. The results give a=1.384, b=$-$0.367 for
$t_2$, and a=1.488, b=$-$0.388 for $t_3$.

There is satisfactory agreement between the three methods. In the
following we take the average of these three determinations as the
value of $M_{WD}$ for the objects without a direct $M_{WD}$ estimate
(see column 13 of Table 1, where the calibrators masses are in
italics).  We adopted as error bars the half difference between the
minimum and maximum values.

We have not used as calibrator the reliable $M_{WD}$ estimate of the
eclipsing object BT Mon ($M_{WD}$=1.05 M$\odot$) because regrettably
and as mentioned above its $t_3$ value is poorly determined.

We recall that the observed white dwarf mass distribution $M_{1}$ in
CVs is in the range 0.8-1.2 M$\odot$, but see Wijnen et al. (2015),
and considerations therein on possible selection effects and on the
contrast with the standard theoretical scenario. In particular, Smith
\& Dhillon (1998) give $0.85\pm 0.05 M\odot$, Nelson et al. (2004)
give $0.95 \pm 0.05$, Savoury et al. (2011) found 0.81 pm 0.04,
Littlefair et al. (2008) give $M_{WD}$ $\sim$ 0.8, Zorotovic et
al. (2011) give 0.83$\pm$ 0.23, while Yuasa et al. (2011) give
$M_{WD}$ $\sim$ 0.88$\pm$0.24 in nonmagnetic CVs.

Instead, the white dwarf mass in classical nova systems, as estimated
by nova frequency, is $\sim$ 1.04-1.24 $M\odot$ (Ritter et al. 1991).
This led Ritter et al. (1991) to conclude that classical novae are a
special sample of CVs, not representative of the intrinsic properties
of CVs. We will return to this in Paper III where we will revisit the
nova paradigm in light of correlations we found among the nova
parameters.

One limitation of our sample of CNe, based as it is on the
  availability of good UV spectra, is that it cannot say anything
  about the highest-mass WDs and their more extreme outcomes (our
  highest WD mass being $1.16\pm0.2$ M$\odot$).

\paragraph{\sl The  radius  of the  white dwarf \\ }
\verb| |

\noindent The radius $R_{WD}$ of the white dwarf enters in the relation between
the disk luminosity and $\dot{M}$. A theoretical nonrelativistic approximation of the relation between $R_{WD}$ and
$M_{WD}$ gives $R_{WD}$ $\sim$ $M_{WD}^{-1/3}$. More accurate
relations between $R_{WD}$ and $M_{WD}$ were proposed by Hamada \&
Salpeter (1961), Nauenberg (1972), Politano et al. (1990), Cannizzo
(1994), Panei et al. (2000), Wade \& Hubeny (1998), and Madej et al.
(2004).

For this work we adopted the Nauenberg (1972) analytical relation,
which is convenient for the calculations of the propagation of the
uncertainties associated  with $M_{WD}$:
 \smallskip

\begin{equation}R_{WD}=0.0115 \cdot (X^{-2/3}-X^{2/3})^{1/2}
\end{equation}

\smallskip

\noindent where X=$M_{WD}$/1.433. This relation agrees well with
previous and recent estimates of $M_{WD}$ and $R_{WD}$.  We have
slightly modified the original constant (0.0112) given by Nauenberg in
order to provide $R_{WD} = 8\cdot10^{-3} R\odot$ in the case of a WD
with $M_{WD}$ = 1 M$\odot$, as suggested by Panei et al. (2000),
Althaus et al. (2013), and Barstow et al. (2015), and Barstow et
al. (2017).

\section {The mass accretion rate  }

The calculation of this parameter, fundamental for our understanding
of CV evolution, requires prior knowledge of the system distance,
the $i$-corrected accretion flux, and the mass and radius of the
primary, that is, the values of all parameters being affected by uncertainties.

In principle, $\dot{M}$ can also be estimated from
a comparison between the observed spectral distribution and that of
synthetic models; see for example Wade \& Hubeny (1998). In this case
one compares the reddening-corrected (far UV) SED with that of
accretion-disk models to derive the $\dot{M}$, the distance, and the
inclination.  However, accretion-disk modeling has not yet reached the
sophistication of stellar-atmosphere modeling and since the number of
parameters in any disk model is rather large, the fitting of the data
to the models does not generally provide unequivocal results.

For this reason, and especially for the availability of  precise
Gaia distances, we preferred an approach based on the
observed UV (and optical) luminosity, after careful treatment
of the various uncertainties and their propagation.

\subsection {$\dot{M}$   from  the  disk
luminosity}

In this section we confidently assume that in the quiescent state
between OBs most of the observed UV and optical luminosity
derives from an accretion disk heated by viscous dissipation of
gravitational energy. Generally, the disk emission of a cataclysmic
variable dominates in the UV decreasing at longer wavelengths.  We
note that CVs have additional radiation sources to the
accretion disks itself: for example the white dwarf, the red dwarf, the hot
spot, and the boundary layer. However in systems with high $\dot{M}$
the disk is the dominating radiation source in the UV and optical
ranges (Patterson 1984; Szkody 2008).

In this case, the estimate of $\dot{M}$ is not model dependent but
requires knowledge of the $i$-corrected disk luminosity
L$_{disk}^{ref}$, and of $M_{WD}$ and $R_{WD}$. If these are known,
$\dot{M}$ can be calculated from the accretion
luminosity:
\begin{equation} \dot{M} = 2\,R_{WD}\,L^{ref}_{disk}\, / \,(GM_{WD})
,\end{equation}
\noindent where it is assumed that half of the gravitational potential
energy of the accreting material is liberated through viscosity in the
accretion disk, the other half being released in the boundary layer
between the innermost disk and the surface layer of the white dwarf
(Prialnik et al. 1989; Frank et al. 2002).

In general, most of the disk accretion luminosity is emitted in the
IUE UV range, while the boundary layer mostly radiates in the EUV and
X-rays. Radiation at wavelengths short of Ly${\alpha}$ is strongly
absorbed and the energy is redistributed to longer wavelengths (Nofar
et al. 1992).

\noindent Numerically, $\dot{M}$ can be represented by
\begin{equation} \dot{M}=5.23 \cdot 10^{-10}\, \phi \,\,
L^{ref}_{disk}/L\odot
,\end{equation}
\noindent where $\phi= 125 \, R_{WD}/M_{WD}$, with radius and mass in
solar units.

The "efficiency" of accretion (i.e., the luminosity associated with a
given $\dot{M}$) is strongly dependent on the compactness of the
accreting object: the higher the ratio $M_{WD}$/$R_{WD}$, the greater
the efficiency. In other words, for a given disk luminosity, $\dot{M}$
will be lower in systems with a massive white dwarf.

The parameter $\phi$ gives the inverse efficiency in converting
$\dot{M}$ into luminosity. Clearly, this efficiency is greater than one in
objects with $M_{WD}$ less than 1 M$\odot$ and lower in objects with
higher $M_{WD}$ values. For a 1-M$\odot$ WD, $R_{WD}$ is $\sim$ 8.0
$\cdot$ 10$^{-3}$ R$\odot$ and $\phi$=1.0.

Columns 6 to 9 of Table 2 give $\dot{M}$ separately for
$M_{WD} = 1 M\odot$ and for individual $M_{WD}$ values.

In particular for V603, HR Del, and RR Pic, the derived value for $\dot{M}$
 can be considered as very reliable because for these three
objects, besides precise Gaia distances, we have access to high-quality UV
spectra with high S/N ratios, accurate extinction correction close to
zero, and reliable estimates of the system inclination.

To compare our $\dot{M}$ values with the averages in the literature we
prefer to use the median of the distribution rather than the mean
because the number of points is not large, and there is no
clear evidence to assume a normal distribution. One advantage of
the
median is that it is a more robust estimator than the mean as it is
less sensitive to "outliers" in the distribution (Mana 2016; Bonamente
2017).

The median of $\dot{M}_{1M\odot}$ is $3.05\cdot 10^{-9}$ M$\odot$
$yr^{-1}$ ($\log \dot{M}_{1M\odot}= -8.52$), while the median of
$\dot{M}_{M_{WD}}$ is $3.29\cdot10^{-9}$ M$\odot$ $yr^{-1}$
($\log \dot{M}_{M_{WD}}= -8.48$).
These median values for $\dot{M}$ are comparable to the average value
($\sim -8.3$) of $\log \dot{M}$ for novae only as given in Fig.  7 of
Patterson (1984). It is worth noting that in that same figure the two
objects with the highest $\dot{M}$ in our sample (HR Del and BT Mon) were
definite "outliers" due to their high $\dot{M}$.

Our median $\dot{M}$ values are lower that the average
$\dot{M} =1.3\cdot 10^{-8}$ M$\odot$ $yr^{-1}$ for quiescent novae
found by Puebla et al (2007) using accretion-disk models. They are
also lower than the value ($\sim 10^{-8}$) found for CVs monitored
immediately before or after a nova OB (Iben et al. 1992). This result
seems to attenuate the need for the hibernation conjecture
(Shara, 1989) to interpret the nova phenomenon. One of the motivations
at the origin of the hibernation idea was the alleged presence, in
post novae, of a disturbingly high $\dot{M}$, of the order of
$5\cdot10^{-8}$ M$\odot$ yr$^{-1}$.

\subsection{Comments on individual objects}

\smallskip \noindent {\bf HR Del.} Hr Del is an outlier in the sample,
with $\dot{M}$ higher by about one order of magnitude
compared to the average value of the other objects in the sample. The
high $\dot{M}$ is close to the value (10$^{-7}$ $\cdot$ M$\odot$
$yr^{-1}$) that would correspond to the steady-burning regime
  (see Fujimoto 1982; Nomoto 1982; Nomoto et al 2007; Iben \& Fujimoto
in Bode \& Evans  2008).  The reason for this continuing activity is not
clear, but
  continuing weak thermonuclear burning may be involved (Friedjung et
  al. 2010).  This peculiar behavior was pointed out by Selvelli \&
Friedjung (2007) who interpreted the presence of a strong P Cyg
profile in the CIV $\lambda$ 1550 resonance doublet as indicative of a
strong steady outflow driven by the high disk luminosity.

\smallskip \noindent {\bf V446 Her.} V446 Her was considered as a
low-mass transfer system (Tappert et al. 2013) and dwarf-nova behavior
was reported by Honeycutt et al (1995).
Our data indicate that its $\dot{M}$ (about 10$^{-9}$ M$\odot$
$yr^{-1}$) is lower than the median, but within the dispersion of the
distribution.  It is surprising that dwarf-nova eruptions, which are
usually interpreted as being a consequence of a thermal instability in a low-$\dot{M}$ accretion disk, may take place also in the presence of the
above-reported $\dot{M}$, a value that is close to that of other
CNe in quiescence.

\smallskip \noindent {\bf GK Per.} Dwarf-nova behavior was reported
also for GK Per (Sabbadin \& Bianchini, 1983). The same arguments as those of
the previous paragraph are valid because $\dot{M}$ in GK Per
($\sim$ 2.0 $\cdot$ 10$^{-9}$ M$\odot$ yr$^{-1}$) is higher than in
V446 Her and close to the average $\dot{M}$ for CNe in quiescence. We
recall that to estimate the $\dot{M}$ of GK Per we used only the IUE
spectra with the lowest flux, that is, those corresponding to the
"quiescent" DN stage.  The  $\dot{M}$ value for GK Per greatly exceeds the
typical mass-transfer rate in DN systems in quiescence and is above
the critical line for thermal instability (Osaki, 1996).  We note that
GK Per, in spectra taken during a DN OB stage, displayed an
increase by a factor of $\sim$70 in the UV luminosity and therefore a
corresponding $\dot{M}$ as high as $\sim$ 1.4 $\cdot$
10$^{-7}$ M$\odot$ yr$^{-1}$.

In conclusion, the observed quiescent $\dot{M}$ of GK Per
(and V446 Her) contradict the basic assumption of the disk-instability
model for the dwarf-nova phenomenon (Osaki, 1996), that is, that
there is no effective accretion from the disk to the WD during
quiescence.

\smallskip \noindent {\bf CP Pup.} For the enigmatic CP Pup
$\dot{M} \sim 1.0 \cdot 10^{-9}$ M$\odot$
yr$^{-1}$ is slightly lower than our median but higher than the one
($\le 1.6 \cdot 10^{-10}$ M$\odot$ yr$^{-1}$) derived by Orio et
al. (2009) via modeling of the X-ray emission only (and assuming a
distance of 1600 pc, higher than the 794 pc from Gaia).  Our
result is in agreement with the indication of Naylor (2002) of high
mass transfer.

The pre-nova magnitude ($\sim 19.4$) was much fainter than the
post-nova one ($\sim 15.0$) (Schaefer \& Collazzi 2010) making CP Pup
the object with the highest pre-post $\Delta m$ ($\sim 5.0$) and
therefore with an expected $\dot{M}$ increase by a factor of about 100
in comparing the values of the pre- and post-OB phases.  Our data,
however, indicate $\dot{M}$ $\sim$ 1.0 $\cdot$
10$^{-9}$ M$\odot$ yr$^{-1}$, close to that of the other ex-novae,
which all returned to the pre-OB magnitude ($\Delta m \sim$ 0)
and have longer orbital periods.

It is worth recalling that White et al.  (1993) and Retter \& Naylor
(2000) suggested that CP Pup is a member of a permanent superhumps
system (Osaki, 1989; Patterson \& Richman, 1991), a new subclass of
CVs with relatively short P and high mass-transfer rates.

\smallskip \noindent {\bf BT Mon.} This object has uncertain OB
parameters: both $m$v$_{max}$ and (consequently) $t_3$.  The nova was
discovered some time after maximum light on Dec 17, 1939, by Wachmann
(1968) and independently by Whipple \& Bok (1946) from plates on
Dec 23, 1939.  On Harvard patrol plates the star was at 8.5 on
Oct 8, 1939, but on Sonneberg plates, Wachmann (1968) found mv=7.6 on
Sept 24, 1939.

Schaefer \& Patterson (1983) discovered five further plates providing
data in OB back to Sep 11, 1939, with an average $m$b$\sim 8.5$, which
they took to be $m$b$_{max}$.  A similar value, $m$v$_{max}$ = 8.1, is
reported in Collazzi et al. (2009).  These values are incompatible
with the Gaia distance of 1412 pc: if $m$v$_{max}$ = 8.1 and
using our reddening E(B-V)=0.24, $M$v$_{max}$ would have a
disturbingly low value of $-3.4$, while novae in OB have an
average $M$v$_{max}= -7.5\pm1.0$; see Sect. 4.

Therefore, $m$v$_{max}$ had to be close to 4.0$\pm$1.0, which would
mean that the earliest observations were taken well after $t_3$.

As an overdue tribute to the work of D. McLaughlin, it must be
remembered that this value confirms his indication (McLaughlin 1941),
based on the spectroscopic evolution and the presence of the emission
lines of NIII $\lambda$ 4640 when the nova was at V=8.5, that the
magnitude at maximum was about four to five magnitudes brighter than V=8.5.
Sanford (1940), based on similar considerations, also speculated that
BT Mon ``may well have been as bright as the third magnitude", while
Smith et al. (1998), based on the spectroscopic considerations of
McLaughlin (1941), also pointed out that 4.2 should have been the
faintest observed magnitude at maximum.

We also incidentally note that any $m$v$_{max} > 5$ would imply an
Eddington ratio lower than one, in contrast with the general behavior of
novae; see Sect. 7. In particular, $m$v$_{max}$ = 8.1 would
give an Eddington ratio of $\sim 6\cdot 10^{-2}$. Based on the
  discussion of $M$v$_{15}$ in Sect. 6, BT Mon should have had
  $m$v$_{15}\sim 6.0$, also brighter than the $m$v$_{max}$ assumed by
  Schaefer (2018).

As mentioned in Sect. 5, Schaefer (2018) included BT Mon in
his "gold" sample and used its apparent low brightness as an
argument against the validity of the MMRD.

Schaefer assumed a low brightness at maximum based on the fact that
"the spectral evidence places the time of maximum around the time of
start of the flat maximum". This is surprising because it is in
contradiction with his and Patterson's 1983 paper, which he uses as a
reference for the statement (see also caption to their Fig 5), where
the "spectral evidence" (from Payne Gaposchkin 1957) was quoted to
say that the magnitude at maximum might have been 3.5 magnitudes
{brighter} than the plateau one he uses now.
All this is also in contrast with the well-documented consideration by
McLaughlin and Sanford reported above on the spectral evolution of BT
Mon.

\section{Correlations (or not) with the accretion rate}

The data contained in Table 1 and Table 2 provide an opportunity to
test the validity of some generally accepted theoretical relations but
also to investigate the possible presence of new ones. Here we outline
some outstanding results that show some unexpected absence and
presence of correlation between $\dot{M}$ and other system parameters.
A comprehensive analysis of all correlations, using standard
statistical procedures, will be presented in Paper III in the
framework of revisiting the nova paradigm.

\begin{figure}
\centering
\resizebox{\hsize}{!}{\includegraphics   [angle=0] {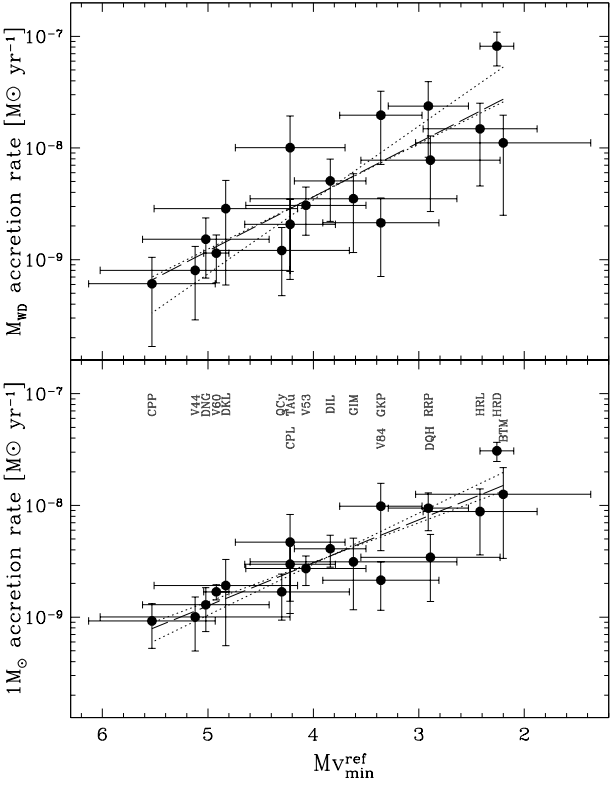}}
\caption{Accretion-rate correlation with the $i$-corrected absolute magnitude at minimum, showing that in the absence
of more information Mv$_{min}^{ref}$ can act as a proxy for $\dot{M}$. {\it Top}: $\dot{M}$ computed using individual WD
masses. {\it Bottom}: $\dot{M}$ computed using $M_{WD}=1M\odot$ for each nova. }
\end{figure}

\subsection{Correlation between $\dot{M}$ and the $i$-corrected
  $M$v$_{min}$ }

It is clear from the results of Paper I and of the sections above that
in old novae $\dot{M}$ is high enough to mask both the WD and
the companion contributions. To quantitatively test this indication we
compared the $\dot{M}$ values derived in Sect. 11 with the
$M$v$_{min}^{ref}$ values derived in Sect. 8.

Figure 3 demonstrates the clear correlation that exists between
$M$v$_{min}^{ref}$ and $\log \dot{M}_{1M\odot}$ and
$\log \dot{M}_{M_{WD}}$. In Fig. 3 (and some other figures showing
$\dot{M}$ in the y-axis), the x-axis increases towards the left. This
is done to maintain similarity with the $\dot{M}$-$t_3$ plot in
Fig. 5.

Linear fits to the data, computed as described in Sect. 2.2, give
\begin{equation} \log \dot{M}_{M_{WD}} = -0.49\pm0.07\,\, M{\rm v}_{min}^{ref} -6.49\pm0.30
,\end{equation}
\begin{equation} \log \dot{M}_{1M\odot} = -0.38\pm0.05\,\, M{\rm v}_{min}^{ref} -6.97\pm0.19
,\end{equation}
\noindent for $\dot{M}$ derived from the individual WD masses and for
a common 1-$M\odot$ WD mass, respectively.

Therefore, in the absence of other spectral information, the
$i$-corrected absolute visual magnitude of old novae and other CVs
accreting at high rates (e.g.,  nova-like) can be used as a convenient
proxy of the actual $\dot{M}$.

It is worth comparing the results derived above with those of other
estimates of $\dot{M}$ from the visual magnitude $M$v.

Lipkin et al. (2001) improved the $\dot{M}$-$M$v relations presented
by Retter \& Leibowitz (1998) and Retter \& Naylor (2000) and derived
the relation
\begin{equation} \dot{M}_{17} = M_{WD}^{-4/3} \, 10^{(5.69 -
0.4M_v^{ref})}
,\end{equation}
\noindent where $\dot{M}_{17}$ is the mass accretion rate in 10$^{17}$
g s$^{-1}$ and $M$v$^{ref}$ is the inclination-corrected absolute
magnitude of the disk. The factor $M_{WD}^{-4/3}$ derives from the
factor $R_{WD}$/$M_{WD}$ in the assumption, generally valid for WDs
with M $\leq$ 1 M$\odot$, that MR$^3$=const. However, this
$R_{WD}$-$M_{WD}$ relation (polytropes) is not accurate in the case of
massive WDs because in this case $R_{WD}$ decreases with a far steeper
slope as $M_{WD}$ increases. We have slightly modified the Lipkin et
al. (2001) relation to account for a more reliable $R_{WD}$/$M_{WD}$
relation based on the Nauenberg $R_{WD}$--$M_{WD}$ formula.

A comparison between our values and those derived from the modified
Lipkin et al. (2001) relation indicates that, on average, the
$\dot{M}_{WD}$ values by Lipkin et al. (2001) are higher by a factor
of about 2.5 $\pm$ 1.4. It should be noted, however, that they made
the implicit assumption that the ratio $L$v$/L_{bol}$ has a
  constant value of $\sim$ 0.14; see also La Dous (1991, 1994) and
Retter \& Leibowitz (1998, 1999).
Instead, our data indicate that the ratio is not constant because it depends on the slope of the SED.

\subsection{No correlation between $\dot{M}$ and $P_{orb}$}

A long-standing and generally accepted paradigm for CVs associates
higher $\dot{M}$ to systems with longer orbital periods (Howell
et al. 2001).  Early studies of the observed $\dot{M}$ in CVs (see
Patterson 1984) indicated a strong dependence of $\dot{M}$ on the
orbital period (all $\dot{M}$ values in this section are in
M$\odot\,{\rm yr}^{-1}$):
\begin{equation} \dot{M} = 5.1\cdot10^{-10} \,\, (P/4)^{3.2}
,\end{equation}
\noindent with $P$ in hours. Patterson (2011) substantially revised
this relation using a larger sample of DNe in OB, confirming the
correlation between Mv (taken as a proxy for $\dot{M}$) and $P$,
although with a weaker linear dependence ($P$ again in hours):
\begin{equation} M{\rm v}_{max}=4.95 - 0.199\,\,P .\end{equation}
\noindent It should be stressed, however, that these results derive
essentially from observations of DNe in OB and have been
somewhat arbitrarily extrapolated to CVs in general and to CNe in
particular. Townsley \& Bildsten (2005) claim that about $50\%$ of CNe
occur in binaries accreting at $\dot{M}\sim10^{-9}$, with $P\sim3-4$
hr, with the remaining $50\%$ split evenly between higher $\dot{M}$
(longer $P$) and lower $\dot{M}$ (shorter $P$); Patterson et
al. (2013) take as round numbers $10^{-8}$ for CNe above the period
gap and $10^{-10}$ for CNe below the gap, while Iben et al. (1992)
consider that the time-averaged mass transfer rate decreases from
$10^{-8}$ at $P\sim6$ hr to $10^{-11}$ at $P\sim80$ min.

\begin{figure}
\centering
\resizebox{\hsize}{!}{\includegraphics   [angle=0] {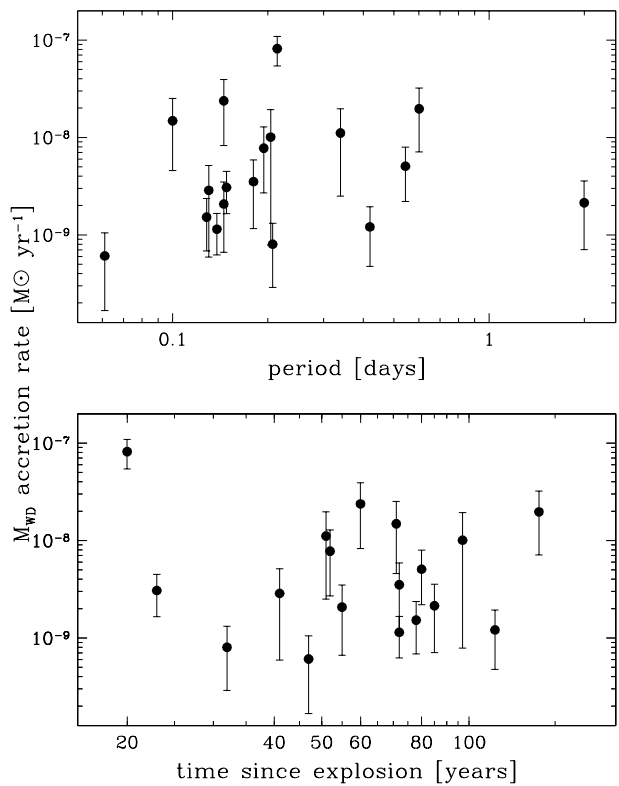}}
\caption{No correlation between the accretion rate and either period
  $P$ or time since explosion $\Delta T$. Plots are in log-log for
  clarity.}
\end{figure}

Our data (see Fig. 4, top) show instead a flat $\dot{M}$ vs. $P$
distribution with a relatively small scatter about a median log value
close to --8.5.

We note that DN Gem and DK Lac fall within the P-gap of 2.15-3.18 hr
(Knigge et al. 2011; Retter et al. 1999), while for HR Lyr there is
some evidence that $P\sim2.4$ hr also inside the gap (Leibowitz et
al. 1995).  However, the $\dot{M}$ values of these three objects are
close to the average of the other old novae of the sample. Also, the
$\dot{M}$ of CP Pup, whose P falls below the gap, is comparable to
that of other novae and higher than that theoretically expected for
CVs below the P-gap (see Patterson 1984; Ritter 2010; Patterson
2013).

The observed flat $\dot{M}$-$P$ distribution of our objects apparently
contradicts the generally accepted paradigm of CVs evolution. We
reiterate, however, that Fig. 7 in Patterson (1984) contains mostly
DNe in OB, with only eight ex-novae, and that these actually show a
flat distribution of $\dot{M}$ (with average $\log\dot{M}$ close to
--8.3). The inclusion of these eight novae has the effect of an
increase in the scatter of data in the upper part of the figure. The
use of a homogeneous class of CVs (e.g., DN only) would have clearly
reduced the scatter.

In conclusion, the ex-novae of our sample that fall above, inside, and
below the P-gap have a median $\log \dot{M}\sim-8.5$ with a nearly
flat distribution with the orbital period P.  This is a challenge to
the generally accepted paradigm of CV evolution and indicates that
CNe indeed represent a special class of CVs, confirming previous
considerations by Ritter et al. (1991) and Livio (1994).

\subsection{No correlation between $\dot{M}$ and $\Delta T$}

Figure 4 (bottom) shows no obvious correlation between $\dot{M}$
and the time elapsed since OB, HR Del being the only recent nova
with high $\dot{M}$.  Due to the fact that our CNe span an interval of
more than one century of nova eruptions, this is a strong indication
that after the explosion  $\dot{M}$ remains
essentially constant during this time interval. This is in agreement
with the results by Weight et al.  (1994), Moyer et al. (2003), and
Thomas \& Naylor (2008) who found no correlation between $\dot{M}$
(or L$_{UV}$) and the time since OB.

The case of the very old nova V841 Oph that has a relatively high
$\dot{M}\sim1.8 \cdot 10^{-8}$ M$\odot\,{\rm yr}^{-1}$ and a
steep continuum in spite of the 140 years elapsed from OB is
remarkable. We note that Engle \& Sion (2005) pointed out that of two UV
spectra of V841 Oph taken 15 years apart, the later one had a higher
flux.

Our data also indicate that if post novae enter a phase of hibernation
(Shara, 1989) $\dot{M}$ does not begin to decline until at
least 150 years post-OB, in agreement with Thomas \& Naylor (2008).
In a recent study of candidate old novae, Tappert et al.  (2015)
pointed out that the two oldest novae in their sample, GR Sgr (Nova
Sgr 1917) and V999 Sgr (Nova Sgr 1910), appear to have the highest
luminosity, contrary to what one would expect from models of nova
evolution.

\begin{figure}
\centering
\resizebox{\hsize}{!}{\includegraphics   [angle=0] {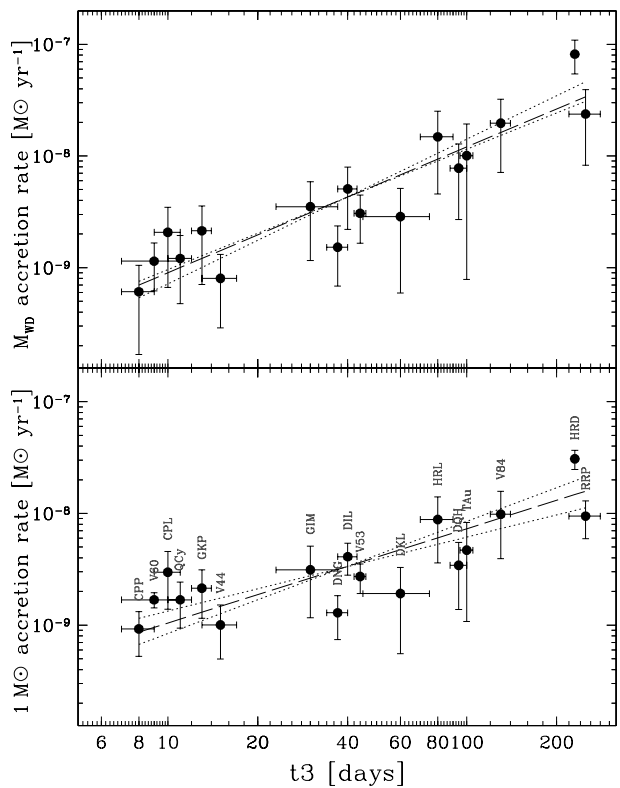}}
\caption{Accretion rate correlation with $t_3$. {\it Top}: $\dot{M}$ computed using individual WD masses.
 {\it Bottom}: $\dot{M}$ computed using $M_{WD}=1M\odot$ for each nova (to show that the correlation is not due to
the possible
degeneracy between $t_3$ and $M_{WD}$).}
\end{figure}

\subsection{Correlation between $\dot{M}$ and the speed class $t_3$}

Our data show that there is a clear correlation between $\dot{M}$
and the speed class $t_3$ (see Fig. 5), indicating that higher
$\dot{M}$ values are associated to objects with larger $t_3$ values.

After linearization, the logs of these two quantities, computed as
described in Sect. 2.2, show the correlations:
\begin{equation}
\log \dot{M}_{M_{WD}}=1.13\pm0.12\,\,\log t_3-10.17\pm0.21
,\end{equation}
\begin{equation}
\log \dot{M}_{1M\odot}=0.84\pm0.12\,\,\log t_3-9.83\pm0.20
,\end{equation}
\noindent for $\dot{M}$ derived from the individual WD masses and for
a common $1M\odot$ WD mass, respectively. We note that via the MMRD
  this implies a correlation between $\dot{M}$ and $M$v$_{max}$ as
  well.

In the past decade models of novae have "weakened" the role of the
WD mass as the dominant parameter (e.g., Townsley and Bildsten 2004, Yaron
et al 2005), showing that for any given mass a large range of
$\dot{M}$ is possible. These in turn determine the ignition mass
(lower for higher accretion) and the temperature at which the nova
OB occurs, with the result that there is a continuum of possible
OB amplitudes, and $t_3$, for a given WD mass.

Within this context the correlation $\dot{M}$-$t_3$ is very
surprising, as it clearly shows a one-to-one interdependence between
$\dot{M}$ and the speed class. This would be
particularly intriguing if $t_3$ turns out to really be a proxy for
the WD mass (as it is often assumed to be), because there are a priori
no obvious reasons why novae with massive WDs should have lower
$\dot{M}$ than novae with lighter WDs. The fact that heavier WD novae
require a smaller total accreted mass to ignite the thermonuclear
explosion (Shara et al. 2010; Glasner \& Truran 2009; Yaron et
al. 2005) does not seem to provide any insight into how this would
control $\dot{M}$ between OBs. Furthermore, while binaries with
massive WD and high $\dot{M}$ may "disappear" from the diagram
because they become common envelope systems (Nomoto 1982), there
appears to be no obvious way to hide the other missing
$M_{WD}$-$\dot{M}$ combinations (unless they become other kinds of
CVs?).

A tempting speculation, if this distribution is confirmed, is that the
strip populated by the old novae discussed here may define the range
of $\dot{M}$ and  $t_3$ values where novae can occur.
Observational $\dot{M}$ determinations for more (old) novae are
clearly needed before exploring this idea any further.

\subsection{Correlation between $\dot{M}$ and OB amplitude}

An interesting correlation exists between the OB amplitude and
$\dot{M}$ (see Fig. 6) that associates large amplitude to low
$\dot{M}$. In hindsight, this is not completely surprising given the
correlation between $M$v$^{icorr}_{min}$ and $\dot{M}$ reported in
Sect. 12.1. Unlike that correlation, which requires knowledge of
several physical parameters (distance, inclination, etc.), this one has
as independent variable: the observed OB amplitude (i.e., the
difference between two apparent magnitudes).

That the correlation exists without the correction for inclination is
somewhat surprising, and may be due to the fact that for most of our
novae the correction is relatively small.  This would also explain why
DQ Her, the nova in the sample with the largest $i$ correction,
appears as an outlier in Fig. 6.

Nevertheless, this correlation provides a {direct} way to
estimate $\dot{M}$ for any nova for which only the OB
amplitude is known. Thus, the OB amplitude acts as a proxy for
$\dot{M}$, much in the way that $t_3$ does (see Sect. 12.4). It
should be noted that Schmidtobreick \& Tappert (2015) suggested that
novae with large OB amplitudes are candidates for low-mass-transfer-rate systems, but they assumed that the absolute magnitude of
a nova explosion differs only slightly for different systems, unlike
what is shown by the L/$L_{\rm Edd}$ and the MMRD relations; see Figs. 1
and 2.

The fits in terms of the explicit observables, computed as described in
Sect. 2.2, are
\begin{equation}
\log \dot{M}_{M_{WD}} = -0.28\pm0.04 \,\, (m{\rm v}_{max}-m{\rm v}_{min}) -5.17\pm0.50
,\end{equation}
\begin{equation}
\log \dot{M}_{M\odot} = -0.21\pm0.03 \,\, (m{\rm v}_{max}-m{\rm v}_{min}) -6.09\pm0.34
.\end{equation}

\begin{figure}
\centering
\resizebox{\hsize}{!}{\includegraphics   [angle=0] {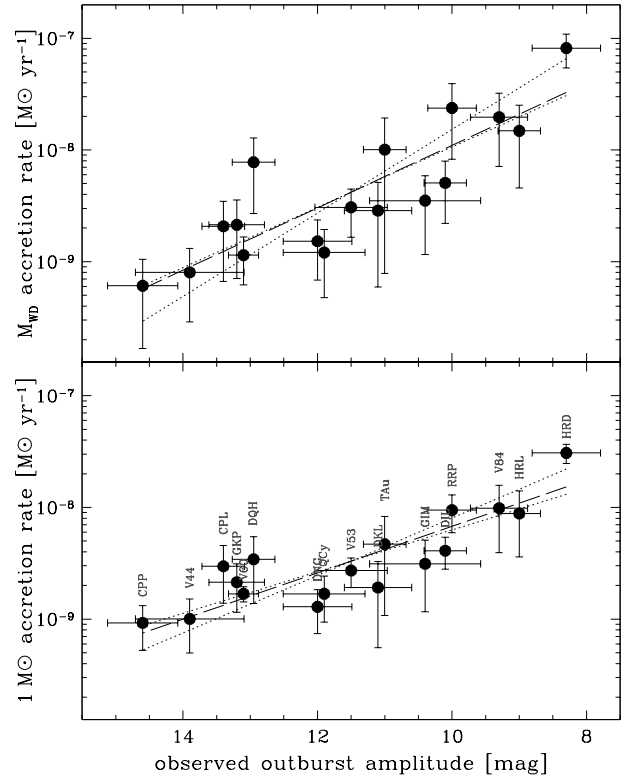}}
\caption{Correlation between accretion rate and the OB amplitude.
{\it Top}: $\dot{M}$ computed using individual WD masses.
{\it Bottom}: $\dot{M}$ computed using $M_{WD}=1M\odot$ for each nova.}
\end{figure}

\section{Concluding summary}

The new and precise distances from Gaia, together with the
data of Paper I and from the literature, allowed us do determine new
system parameters for our sample of 18 old novae with UV data
({e.g.,} $M$v$_{max}$, $M$v$_{min}^{\rm i-corr}$,
$L_{OB}/L_{\rm Edd}$, $L_{\rm disk}^{\rm i-corr}$, $\dot{M}$).

The final values include a detailed treatment of the errors and
their propagation and are summarized in Table 2. Functional relations
are provided in the text for all correlations (either revisited or
new) presented in the paper.
Our main results can be summarized as follows.

\begin{enumerate}

\item

The average visual magnitude at maximum for the novae in the sample
is $-7.5\pm1.0$.

\smallskip \item

The bolometric luminosity of all 18 objects at maximum is  equal to or
above the Eddington limit, with an Eddington ratio in the range $\sim$1
to
$\sim$20 for $L_{Edd}(1M\odot)$, and $\sim$1 to $\sim$17 for
  $L_{Edd}(M_{WD})$.

\smallskip \item

Various parameters correlate with the speed class $t_3$, for example
$M$v$_{max}$ (the MMRD relation) and $L_{bol}^{max}/L_{Edd}$, showing
that the brightest and most super-Eddington novae correspond to the
shortest $t_3$values.

These relations are (or were) also useful to derive other parameters
(e.g., the distance from the MMRD since knowing $t_3$ provides an
estimate of the absolute magnitude at max).

\smallskip \item

The median $\dot{M}$ for the 18 old novae is
$\log \dot{M}=-8.52$ for M$_{WD}\equiv1$M$\odot$, and
$\log \dot{M}= -8.48$ for individual M$_{WD}$ values. These results
are not model dependent and are essentially based on the (reddening-corrected) UV luminosity after correction for inclination effects.

\smallskip \item

The $\dot{M}$ values are lower than those so far assumed in the
studies of CNe evolution, and appear to attenuate the need for a
hibernation conjecture to interpret the nova phenomenon.

\smallskip \item

There is a surprising correlation between $\dot{M}$ and
the speed class $t_3$ ($\dot{M}$ increasing with $t_3$). We cannot
find a simple explanation that could account for this. How, for example,
does $t_3$, a likely proxy for the WD mass, control $\dot{M}$
between OBs? Clearly this requires a more in-depth theoretical
analysis.
Yet, this is a useful relation that allows one to estimate $\dot{M}$
directly, and marks $t_3$ as an extremely convenient observable to
evaluate several critical parameters of classical novae.

\smallskip \item

The average $i$-corrected absolute visual magnitude at minimum of the
old novae in our sample is 3.9$\pm$1.0. This parameter also correlates
with $\dot{M,}$ meaning that, in the absence of other spectral information, it
can be used as a convenient proxy for $\dot{M}$ in
old novae and other CVs accreting at high rates (e.g., nova-like).

\smallskip \item

Another useful correlation is the one between the OB amplitude
and $\dot{M}$. The advantage here is that the difference
between two apparent magnitudes provides a {direct} way to
estimate $\dot{M}$ for any nova for which only the OB
amplitude is known.

\smallskip \item

There is {no} apparent correlation between $\dot{M}$ and the time
elapsed from OB. Due to the fact that our CNe span an interval
of more than one century of nova eruptions, this indicates that after
the explosion $\dot{M}$ remains essentially
constant during this time interval. This also indicates that if ex-novae enter a phase of hibernation (Shara, 1989) $\dot{M}$
does not begin to decline until at least 150 years after OB.

\smallskip \item

In contrast with the commonly accepted paradigm of CV evolution based
on the disrupted magnetic braking scenario (Rappaport et al. 1983;
Howell et al. 2001) our data show a flat distribution with {no}
correlation between $\dot{M}$ and the orbital period $P$. This
suggests that novae are a special class of CVs, and confirms previous
considerations by Ritter (1991) and Livio (1992c, 1994); see also Zorotovic
et al. (2016).

\smallskip \item

GK Per and V446 Her, the two ex-novae in our sample that have
undergone DN eruptions, have a rather high $\dot{M}$ ($\sim$ 1-2
$\cdot$ 10$^{-9}$ M$\odot$ yr$^{-1}$) during the DN quiescent phase,
similar to that of the other remnants. This $\dot{M}$ value is well
above the critical line for thermal instability, (Osaki, 1996) and is in
severe contradiction with the basic assumption in the disk instability
model of no effective accretion from the disk to the WD during
quiescence.

\end{enumerate}

\noindent It is possible that all the correlations are an indication
that the quantities we have considered depend on the same hidden
parameter. This can be traced back to the WD mass: since $\dot{M}$
correlates with $t_3$, and $t_3$ is a likely proxy for $M_{WD}$,
the  correlation of a
parameter with $\dot{M}$ may imply a correlation with
$M_{WD}$.  This would not be surprising, as $M_{WD}$ is considered a
dominant parameter in the theory of novae.

The range of fit coefficients of the various correlations may in fact
help us to better understand the dependences between physical
quantities. Indeed, the correlations may possibly even help clarify
the physics of the nova phenomenon. We will further pursue this line
of reasoning in Paper III of this series.

\begin{acknowledgements} We wish to dedicate this paper to the memory
  of our friends and colleagues Angelo Cassatella and Michael
  Friedjung, with whom we shared inspiring discussions
in the early
  stages of this study.  We gratefully thank Elena Mason and Bob
    Williams for useful discussions, Carlo Morossi, Maria Grazia
  Franchini and Massimo Della Valle for fruitful comments, and
  Jorge Melnick and Jason Spyromilio for constructive debates on the
  intricacies of linear regressions. We thank the anonymous
    referee for catching a mistake in the original version, and for
    the useful comments and suggestions that helped us improve the
    paper.
\end{acknowledgements}

\appendix
\section{The  propagation  of errors. }

We use here the term error to refer to the uncertainty in the value of
a variable (this is what we have so-far referred to as a nova parameter, or a
physical quantity).  Propagation of errors is essential for
understanding how the uncertainty in a variable affects the
computations that use that variable.  A basic assumption in the theory
of error propagation is that that the individual variables are
uncorrelated and independent and that the errors are symmetric
(Gaussian), see Barlow (1989), Bevington $\&$ Robinson (1992), Taylor
(1997).  In this case, in the simple case of a sum (or product) of two
quantities the errors (the relative errors) are added in quadrature.
In the more general case of a complex function one has to compute the
total derivative of that function, a task that, in some cases may be
demanding.

As mentioned in Sect. 2, an important aspect of this study of novae
derives from the effort we have made to estimate the errors associated
to the basic physical quantities.  The calculation of the error
propagation up to the values of the most wanted parameters, for example  the
accretion disk luminosity and the accretion rate is a relatively onerous
exercise. As an illustration of this we note that if one considers
explicitly all the relevant parameters, the final expression for $\dot{M}$
(eq. 10) can be written as
\begin{equation} \dot{M}=1.304 \cdot 10^{-7} \, (R_{WD}/M_{WD}) \, d^2
\int_{1100}^{6000} A \, \lambda^{-\alpha} \,
f^{-1}_{i,\lambda}\, d\lambda
.\end{equation}
It is clear from the above equation that the final value of $\dot{M}$
critically depends on quantities like the $i$-corrected or reference
integrated flux, the distance, the white dwarf mass, and the white
dwarf radius. In turn, the reference integrated flux depends on the
color excess E(B-V) that determines the constant A and the index
$\alpha$ of the PL approximation of the UV SED, and the
$f^{-1}_{i,\lambda}$ factor that takes into account both the
geometrical and the limb-darkening corrections for the inclination of
the system. The values of these individual parameters are all affected
by uncertainties whose propagation up to the final product, the disk
luminosity or $\dot{M}$, must be correctly evaluated.  To do
this we utilized two separate methods: the standard pencil and paper
calculation of partial derivatives and their sum, and the use of the
Python (Anaconda) environment that provides specific modules and
packages (numpy: see www.numpy.org, and unumpy-uncertainties: see
http://pythonhosted.org/uncertainties/) that allow a less cumbersome
but ``black box'' calculation of complex operations in arrays with
errors. Both methods yielded the same results within less than
1\%. The numbers in the tables are from the Python output.

A special case is that of the power law approximation to the SED. The
uncertainty in the color excess E(B-V) around the central value gives
upper and lower limits of the index $\alpha$ and of the corresponding
constant A of the PL approximation.  Since alpha and A are  not
independent, we calculated the central value and the upper and lower
limits of the integrated flux for each of the three pairs of
($\alpha$, A). These integrals are slightly asymmetric so we have
taken as error for the PL integral the semi-difference between the
upper and lower value of the lambda integrated flux. In this way we
could quadratically combine them with the relative errors determined
by the uncertainties in the inclination for every ($\alpha$, A)
pairs. This quadratic sum gives, for each star, the total error in
what we define as the reference integrated flux due to the combined
uncertainties in E(B-V) and the inclination. Finally the reference
integrated flux with its error has been combined with the remaining,
independent parameters of eq A1, that is, d, $M_{WD}$, $R_{WD}$, to
determine the final values, with errors, of $\dot{M}$.

Of course, for Paper I we also explored fitting power law and
reddening curve simultaneously to the UV flux. This would have had the
advantage of providing symmetric fitting errors from the matrix
inversion. However this method is very sensitive to the way emission
and absorption lines (and in the case of IUE noise spikes) are either
avoided or removed from the spectrum so that we felt more confident
about the error determination through the more interactive method
above.

The error propagation for all other final parameters (\emph{e.g.,}
$L_{Edd}$, $i$-corrected absolute magnitude), has been estimated much
more easily because the associated parameters (e.g.,  E(B-V), distance,
$m$v$_{max}$, $m$v$_{min}$) are independent of each other.

\end{document}